\documentclass[12pt]{article}
\RequirePackage{amsthm,amsmath,amsfonts,amssymb,booktabs,tabularx,array,algorithm,algpseudocode,subcaption}
\RequirePackage[authoryear,round]{natbib}
\RequirePackage[colorlinks,citecolor=blue,urlcolor=blue]{hyperref}
\RequirePackage{graphicx,xcolor}
\theoremstyle{plain}
\newtheorem{theorem}{Theorem}[section]

\newtheorem{lemma}{Lemma}

\theoremstyle{definition}
\newtheorem{remark}{Remark}

\newtheorem{assumption}{Assumption}

\title{Graph-Enabled Efficient Federated Bayesian Modeling}

\author{Chenyang Zhong\thanks{Corresponding authors}\\
     Department of Industrial Engineering and Operations Research,\\
     University of California, Berkeley\\
     \url{cyzhong@berkeley.edu}
     \and 
     Shouxuan Ji \\
     Department of Statistics and Data Science,\\
     National University of Singapore\\
     \url{shouxuan.ji@u.nus.edu}
     \and
     Tian Zheng\footnotemark[1]\\
     Department of Statistics, Columbia University\\
     \url{tian.zheng@columbia.edu}
     }

\begin{document}

\maketitle

\vspace{-2em}

\begin{abstract}
Federated Bayesian modeling requires combining evidence across distributed data holders while preserving posterior uncertainty and keeping local data decentralized. We propose Federated Latent Graph MCMC (FLaG-MCMC), a framework for sequential posterior transfer in which accumulated posterior information is represented by a reservoir of Monte Carlo samples. FLaG-MCMC constructs a graph that captures the geometry of the posterior reservoir and uses graph-enabled proposals whose per-iteration cost is independent of the reservoir size. Each data holder updates the inherited posterior using its local likelihood and returns an updated reservoir for subsequent inference. 

We apply FLaG-MCMC in two health science settings. Using mobile and wearable device data from GLOBEM, we investigate associations between depression status and behavioral patterns while sequentially accumulating posterior information across participants. FLaG-MCMC accurately reconstructs the corresponding pooled Bayesian posterior while closely tracking evolving posterior location and uncertainty in these associations. We also conduct a simulation study motivated by Bayesian evidence synthesis for opioid use disorder prevalence across studies, transferring posterior information from an Ohio study to a subsequent New York analysis under covariate shift. In this setting, FLaG-MCMC demonstrates reliable posterior transfer and favorable computational scaling. Theoretical results establish the validity and consistency of the framework.
\end{abstract}

\section{Introduction}
\label{sec:introduction}

Modern data-driven research in statistics and AI increasingly requires integrating evidence across distributed, heterogeneous, and sensitive data sources. In federated learning, observations remain with individual users or institutions while information is communicated for global learning \citep{mcmahan2017communication,li2020federatedlearning,kairouz2021advances}; analogous settings arise in meta-analysis, where evidence from separate studies must be synthesized, and in transfer learning, where information from a source population is leveraged for inference in a related target population \citep{borenstein2021introduction,suder2023bayesian}. In many scientific applications, uncertainty quantification is essential for assessing the strength of evidence, characterizing heterogeneity across data sources, and supporting subsequent inference and decision-making. Bayesian inference provides a natural framework for these settings by representing accumulated
information through a posterior distribution that captures parameter uncertainty. Sequential Bayesian updating---in which the posterior from one analysis serves as the prior for the next---further provides a principled mechanism for propagating this information across studies, participants, or populations \citep{wurpts2021sequential,suder2023bayesian}.

A central challenge is how to transfer and subsequently update a posterior distribution when it is available only through Monte Carlo samples rather than a tractable density. This situation arises naturally when posterior draws summarize an earlier Bayesian analysis, encode information learned from a related population, or represent posterior information accumulated from preceding participants in a distributed study. 

The need for such posterior transfer is particularly salient in mobile health, where smartphones and wearable devices generate rich longitudinal measurements of individual behavior while participant-level observations
naturally remain distributed. Our primary real data application analyzes GLOBEM \citep{xu2022globem}, a multi-year mobile health resource combining passive smartphone and wearable sensing with survey-based measures of well-being. We study associations between depression status and interpretable behavioral summaries of mobility, phone usage, physical activity, and sleep. From a federated Bayesian perspective, each participant contributes local evidence about shared behavioral associations, with posterior uncertainty accumulated sequentially across participants. Successive participants can provide heterogeneous information, and the posterior location and uncertainty may evolve materially. 

To carry out these sequential updates from posterior draws, a natural approach is to use kernel density estimation (KDE) \citep{rosenblatt1956remarks,parzen1962estimation} to construct a continuous representation that can serve as the prior for the next Bayesian update. Combining this representation with the current data holder's likelihood yields the updated posterior. We consider both ordinary kernel representations and, for longer sequences of
posterior transfer, a moment-preserving shrinkage construction building on the shrinkage principle of \citet{west1993approximating} and \citet{liu2001combined}.

Sampling from the resulting posterior, however, introduces a key computational bottleneck. A reservoir containing $B$ posterior draws induces a $B$-component kernel representation of the prior, and standard Metropolis--Hastings sampling requires $O(B)$ kernel calculations per iteration. Increasing $B$ can improve the fidelity of posterior representation while simultaneously increasing the computational cost of posterior sampling. Efficient posterior transfer thus requires a sampling strategy that can exploit a large posterior reservoir without repeatedly evaluating the complete mixture.

To address this challenge, we introduce Federated Latent Graph MCMC (FLaG-MCMC), an efficient graph-enabled Markov chain Monte Carlo method for sequential posterior transfer in federated settings. FLaG-MCMC constructs a graph that captures the geometry of the posterior reservoir and augments the posterior with an anchor variable representing a graph node. The resulting Markov chain combines local moves among neighboring graph nodes with global refreshes that permit movement between distant regions of the reservoir. Each transition requires evaluating only the kernel components associated with the current and proposed anchors, yielding per-transition computational cost independent of $B$ while preserving the intended posterior target.

In the sequential federated procedure, a central server maintains the posterior reservoir representing information accumulated through preceding data holders. Each new data holder uses this inherited reservoir together with its local likelihood to perform the next Bayesian update and returns the resulting posterior draws to the server. Only posterior draws of shared global parameters are communicated, while local observations and holder-specific latent variables remain with the corresponding data holder.

Applied to GLOBEM, FLaG-MCMC sequentially accumulates posterior information across participants while accurately reconstructing the corresponding pooled Bayesian posterior at each stage. In particular, it closely tracks changes in the location and uncertainty of the posterior distributions for the behavioral associations. Several associations exhibit substantial posterior evolution across participants and study years, illustrating the inferential value of retaining the posterior distribution throughout the sequential federated analysis.

A complementary healthcare setting arises when Bayesian evidence needs to be synthesized across studies or populations. We investigate this setting through a simulation study motivated by Bayesian estimation of opioid use disorder (OUD) prevalence \citep{doogan2022opioid}. Estimating the prevalence of untreated OUD is particularly challenging because relevant information may be distributed across health systems and jurisdictions, while restrictions on individual-level data sharing often preclude direct data pooling. We consider posterior transfer from an Ohio study to a subsequent New York analysis under substantial covariate shift, with the Ohio posterior communicated through Monte Carlo draws rather than the underlying observations. FLaG-MCMC closely reconstructs the pooled Bayesian posterior while achieving substantially lower computational cost compared with a Metropolis random walk. As the reservoir size increases, the per-iteration cost of the Metropolis random walk grows approximately linearly, whereas that of
FLaG-MCMC remains nearly constant. Together, these results demonstrate that FLaG-MCMC provides reliable posterior transfer across studies with favorable computational scaling.

We complement these applications with theoretical analysis of the posterior approximation and the graph-enabled Markov chain. We first show that, at each stage, given the finite posterior reservoir, the graph-enabled Markov chain has the intended kernel-based posterior as its stationary distribution. We then establish the consistency of sequential posterior transfer, showing that the kernel-based posterior converges to the corresponding pooled oracle posterior at each stage under appropriate regularity conditions. Finally, we derive an explicit mixing time bound for the graph-enabled Markov chain.

\smallskip

\noindent\textbf{Contributions.}
Our contributions are threefold. First, we develop FLaG-MCMC, a framework for sequential Bayesian posterior transfer from Monte Carlo draws that combines flexible kernel representations with graph-enabled sampling. The resulting sampler has per-transition computational cost independent of the reservoir size while preserving the intended posterior target. The framework accommodates both ordinary kernel representations and a moment-preserving shrinkage construction for repeated posterior transfer, as well as models with holder-specific latent variables that remain local. Second, we establish theoretical properties of FLaG-MCMC, showing that the graph-enabled Markov chain has the intended kernel-based posterior as its stationary distribution, establishing the consistency of sequential posterior transfer with respect to the pooled oracle posterior, and deriving an explicit mixing time bound. Third, we apply FLaG-MCMC
to two complementary health science settings. Using mobile and wearable sensing data from GLOBEM, we conduct participant-level federated Bayesian analysis of depression, sequentially accumulating posterior information while tracking evolving uncertainty in behavioral associations. In a simulation study motivated by OUD prevalence estimation, we study posterior transfer across populations under covariate shift and demonstrate accurate reconstruction of pooled Bayesian inference together with favorable computational scaling.

\smallskip

\noindent\textbf{Related work.}
Federated learning provides a framework for collaborative model training across decentralized devices and data sources, with applications spanning a wide range of scientific and technological domains \citep{mcmahan2017communication,li2020federatedlearning,dayan2021federated,kairouz2021advances,zhou2024federated,cai2025cost,mangold2025refined,yin2026distribution}. Bayesian approaches to federated learning formulate this problem as posterior inference over shared parameters, with methods based on posterior averaging \citep{al2021federated}, variational inference and expectation propagation \citep{guo2023federated,kim2025fedhb}, and piecewise deterministic Markov processes \citep{bierkens2022federated}, among others; see \citet{cao2023bayesian} and the references therein for a comprehensive overview. Earlier work on distributed Bayesian computation combines subset posteriors through averaging \citep{scott2016bayes}, density
product estimation \citep{neiswanger2014asymptotically}, or Wasserstein barycenters \citep{srivastava2018scalable}. These approaches typically construct subset posteriors separately and subsequently aggregate them to approximate the full-data posterior. Our setting instead considers sequential Bayesian updating, where the posterior accumulated from preceding data holders is represented nonparametrically through a reservoir of Monte Carlo draws and evolves as each new data holder contributes information. We represent this inherited posterior continuously through a kernel mixture and use graph-enabled transitions to update it efficiently at each stage. We note that our setting---in which the likelihood is available but the prior is known only through samples---is distinct from likelihood-free inference \citep{beaumont2002approximate}, where the prior is known but the likelihood is intractable. The use of historical posterior information connects more broadly to power priors \citep{ibrahim2000power} and Bayesian transfer learning \citep{xuan2021bayesian,suder2023bayesian}.

\smallskip

\noindent\textbf{Organization.}
Section~\ref{sec:method} introduces FLaG-MCMC, including the sequential posterior transfer setup, kernel representation of the posterior reservoir, and graph-enabled posterior updating. Section~\ref{sec:theory} establishes theoretical properties of the proposed framework. Section~\ref{sec:exp-oud} presents the OUD simulation study, and
Section~\ref{sec:globem} presents the GLOBEM application. Section~\ref{sec:discussion} concludes with a discussion of further implications and future directions.

\section{Federated Latent Graph MCMC}
\label{sec:method}

This section introduces our core methodology, Federated Latent Graph MCMC
(FLaG-MCMC), for sequential Bayesian posterior transfer across distributed
data holders. We formulate the federated Bayesian updating problem, introduce kernel representations of the transmitted posterior reservoir, and develop the latent graph construction and the graph-enabled MCMC transition. We conclude with the sequential FLaG-MCMC procedure and its extension to models with local latent variables.

\subsection{Federated Bayesian Posterior Transfer}
\label{sec:posterior-transfer}

Consider $N$ data holders arriving sequentially, where data holder $i$
observes local data $\mathcal D_i$. Let
$\theta\in\Theta\subseteq\mathbb R^d$ denote the global parameter shared
across data holders, and let $L_i(\theta)$ denote the likelihood contribution
from $\mathcal D_i$. When local latent variables are present and can be
marginalized, $L_i(\theta)$ denotes the corresponding marginal likelihood;
Section~\ref{sec:local-latent} considers the more general setting in which
local latent variables are retained and updated locally.

Let $\pi_0(\theta)$ denote the initial prior for the global parameter. If
Bayesian analyses of historical data or related studies are available, $\pi_0$ can be taken as the resulting posterior distribution, typically represented
only through Monte Carlo draws. If a pretrained model is available---for
example, a foundation language model from which topic structures can be
extracted \citep{bommasani2021opportunities}---$\pi_0$ may incorporate this
learned knowledge as an informative prior. In the absence of either source,
$\pi_0$ reduces to a standard model prior. After incorporating the first
$i$ data holders, the pooled Bayesian posterior is
\begin{align*}
\pi_i(\theta)
&\propto
\pi_0(\theta)
\prod_{j=1}^{i}
L_j(\theta).
\end{align*}

Under the federated constraint, each data holder retains its local data and
communicates only posterior information about the global parameter to a
central server. Sequential Bayesian updating gives $\pi_i(\theta)\propto  \pi_{i-1}(\theta) L_i(\theta)$, but the preceding posterior is generally available to data holder $i$ only through a finite
reservoir of draws maintained by the server. Denote this reservoir by $\mathcal R_{i-1}=\{\theta_b^{(i-1)}\}_{b=1}^{B}$. To facilitate the next Bayesian update, we represent the reservoir by the continuous kernel mixture
\begin{align*}
\widetilde\pi_{i-1}(\theta)
:=
\frac{1}{B}
\sum_{b=1}^{B}
K_h
\left(
\theta-m_b^{(i-1)}
\right),
\end{align*}
where $K_{h}$ is a kernel with bandwidth $h>0$ and $m_b^{(i-1)}$ denotes the center of component $b$. Section~\ref{sec:kernel-representation} introduces the ordinary and moment-preserving choices for these component centers. Data holder $i$ then
targets $\widehat\pi_i(\theta)\propto
\widetilde\pi_{i-1}(\theta) L_i(\theta)$
using only its local data and the transmitted posterior representation.

After the update, posterior draws of $\theta$ are returned to the central
server to form the reservoir for the next data holder, while $\mathcal D_i$ remains local. This sequential posterior transfer mechanism propagates accumulated information without requiring raw data to be shared across data holders.

\subsection{Kernel Representation of the Posterior Reservoir}
\label{sec:kernel-representation}

The kernel representation determines how a finite posterior reservoir is converted into a continuous distribution for subsequent Bayesian updating. We consider two variants. Ordinary kernel density estimation (KDE) provides a direct nonparametric representation of the posterior reservoir. For posterior transfer across many data holders, we additionally consider a moment-preserving shrinkage representation that preserves the first two empirical moments of the transmitted reservoir.

\subsubsection{Ordinary Kernel Representation}\label{sec:ordinary-kernel}

Temporarily suppressing the data-holder index, let $\{\theta_b\}_{b=1}^{B}$ denote the posterior reservoir. Let $K:\mathbb R^d\to[0,\infty)$ be a kernel satisfying $\int_{\mathbb R^d}K(u)\,du=1$. For a bandwidth $h>0$, define
\begin{align*}
    K_h(u):=h^{-d}K(h^{-1}u).
\end{align*}
The ordinary kernel representation is
\begin{align*}
\widetilde\pi^{\mathrm{KDE}}(\theta)
:=
\frac{1}{B}
\sum_{b=1}^{B}
K_h(\theta-\theta_b).
\end{align*}
Thus, in the general kernel representation introduced in Section~\ref{sec:posterior-transfer}, the component centers are
$m_b=\theta_b$, $b=1,\ldots,B$.

\subsubsection{Moment-Preserving Shrinkage Representation}
\label{sec:shrinkage}

Let $\bar\theta$ and $S$ denote
the empirical mean and covariance of the posterior reservoir,
\begin{align*}
\bar\theta
:=
\frac{1}{B}
\sum_{b=1}^{B}
\theta_b, \qquad
S
:=
\frac{1}{B}
\sum_{b=1}^{B}
(\theta_b-\bar\theta)
(\theta_b-\bar\theta)^\top.
\end{align*}
For the shrinkage construction, we additionally assume that $K$ is centered ($\int_{\mathbb R^d}uK(u)\,du=0$) and has finite second moments, with covariance matrix $\Omega_K
:=
\int_{\mathbb R^d}uu^\top K(u)\,du$. The ordinary
kernel representation then satisfies
\begin{align*}
E_{\widetilde\pi^{\mathrm{KDE}}}(\theta)
=
\bar\theta,\qquad
\operatorname{Cov}_{\widetilde\pi^{\mathrm{KDE}}}(\theta)
=
S+h^2\Omega_K.
\end{align*}
Thus ordinary KDE preserves the empirical mean, while its covariance includes
the additional kernel covariance $h^2\Omega_K$. For longer sequences of posterior transfer, it can be desirable to preserve the first two empirical moments of each transmitted reservoir. We therefore
shrink the kernel centers toward the empirical posterior mean while retaining
smoothing, following the shrinkage principle used in kernel posterior approximation and sequential Monte Carlo
\citep{west1993approximating,liu2001combined}.

Let $m_b=\bar\theta+A(\theta_b-\bar\theta)$, where
$A\in\mathbb R^{d\times d}$ is chosen to satisfy $ASA^\top+h^2\Omega_K=S$. The resulting shrinkage representation is
\begin{align*}
\widetilde\pi^{\mathrm{sh}}(\theta)
:=
\frac{1}{B}
\sum_{b=1}^{B}
K_h
\left(
\theta-\bar\theta-A(\theta_b-\bar\theta)
\right).
\end{align*}
Because the transformed component centers $\{m_b\}_{b=1}^{B}$ average to $\bar\theta$,
\begin{align*}
E_{\widetilde\pi^{\mathrm{sh}}}(\theta)
=
\bar\theta,\qquad
\operatorname{Cov}_{\widetilde\pi^{\mathrm{sh}}}(\theta)=ASA^\top+h^2\Omega_K
=S.
\end{align*}
The shrinkage representation therefore preserves the first two empirical moments
of the transmitted reservoir exactly.

A convenient implementation is obtained by whitening the posterior reservoir.
Let $S=LL^\top$ and define $z_b:=L^{-1}(\theta_b-\bar\theta)$, so that the
whitened reservoir has empirical covariance $I_d$. Let $\widetilde{K}$ be a centered
kernel with covariance $I_d$ in the whitened coordinates. Transforming $\widetilde K$ back to the original coordinates gives the centered kernel
$K(u):=|\det(L)|^{-1}\widetilde K(L^{-1}u)$, with covariance
$\Omega_K=S$. For $h\in (0,1]$, taking $A=\sqrt{1-h^2}I_d$ gives the component centers
\begin{align*}
m_b
=
\bar\theta
+
\sqrt{1-h^2}
(\theta_b-\bar\theta).
\end{align*}
The resulting shrinkage representation is therefore
\begin{align*}
\widetilde\pi^{\mathrm{sh}}(\theta)
=
\frac{1}{B}
\sum_{b=1}^{B}
K_h(\theta-m_b).
\end{align*}

The ordinary and moment-preserving constructions provide two kernel representations within the same FLaG-MCMC framework. The graph-enabled MCMC
transition described next applies to either representation.

\subsection{Latent Graph and Graph-Enabled MCMC Transition}
\label{sec:graph-mcmc}

A large posterior reservoir provides a flexible representation of the inherited posterior, but direct Metropolis--Hastings sampling from the resulting kernel-mixture posterior
\begin{align}\label{poster_pi}
\widehat\pi_i(\theta)
&\propto
\left(
\frac{1}{B}
\sum_{b=1}^{B}
K_h(\theta-m_b)
\right)
L_i(\theta)
\end{align}
is computationally expensive because it requires evaluating all $B$ kernel components at every proposed state. FLaG-MCMC avoids this repeated evaluation of the full mixture by introducing an auxiliary anchor variable that indexes a kernel component and using the geometric structure of the posterior reservoir to guide transitions between anchors.

To capture this geometry, FLaG-MCMC constructs a neighborhood graph $G$
over the $B$ kernel components. The graph can be built using an exact $k$-nearest-neighbor graph, approximate $k$-nearest-neighbor methods such as HNSW \citep{malkov2018efficient} or FAISS \citep{johnson2019billion}, or a $k$-nearest-neighbor graph after projection of the posterior reservoir to a lower-dimensional subspace. The graph guides efficient exploration of the posterior target, while the target itself is determined by the kernel representation together with the current data holder's likelihood.

The graph $G$ enters the sampler through an auxiliary anchor variable $b\in\{1,\ldots,B\}$ identifying a kernel component. At data holder $i$,
define the augmented target
\begin{align}\label{pipostetheta}
\widehat \Pi_i(\theta,b)
&\propto
K_h(\theta-m_b)
L_i(\theta).
\end{align}
Marginalizing over the anchor gives
\begin{align*}
\sum_{b=1}^{B}
\widehat \Pi_i(\theta,b)
&\propto
\frac{1}{B}
\sum_{b=1}^{B}
K_h(\theta-m_b)
L_i(\theta)
\propto
\widehat\pi_i(\theta),
\end{align*}
and therefore recovers the target posterior.

Let $\mathcal N(b)$ denote the set of neighbors of anchor $b$ in $G$. Fix $\rho\in (0,1)$. The anchor proposal combines graph-guided local moves with a uniform global refresh,
\begin{align*}
q_G(b,b')
&:=
\rho\cdot\frac{1}{B}
+
(1-\rho)\cdot
\frac{
I\{b'\in\mathcal N(b)\}
}{
|\mathcal N(b)|
},
\end{align*}
where $I\{\cdot\}$ denotes the indicator function. The local component proposes among neighboring graph nodes, which tend to yield moderate likelihood ratios; the global refresh permits direct movement between distant regions of the reservoir, allowing the Markov chain to escape local basins.

Given the current state $(\theta,b)$, propose $b'\sim q_G(b,\cdot)$ and then
$\theta'\sim Q_\sigma(\cdot\mid m_{b'})$, where $Q_\sigma(\cdot\mid m)$ is a proposal kernel centered at $m$ with scale
parameter $\sigma$ (for example, $N(m, \sigma^2 I_d)$). The Metropolis--Hastings acceptance probability is
\begin{align}
\alpha\big((\theta,b),(\theta',b')\big)
=
\min\left\{
1,
\frac{
K_h(\theta'-m_{b'})
L_i(\theta')
q_G(b',b)
Q_\sigma(\theta\mid m_b)
}{
K_h(\theta-m_b)
L_i(\theta)
q_G(b,b')
Q_\sigma(\theta'\mid m_{b'})
}
\right\}.
\label{eq:flag-mh}
\end{align}
The proposed state $(\theta',b')$ is accepted with probability $\alpha\big((\theta,b),(\theta',b')\big)$; otherwise, the chain remains at
$(\theta,b)$. Iterating this transition yields an MCMC sampler targeting
$\widehat\Pi_i(\theta,b)$, whose marginal distribution in $\theta$ is
$\widehat\pi_i(\theta)$.

Crucially, each transition requires evaluating only the kernel components
associated with the current and proposed anchors. The complete $B$-component mixture does not enter the acceptance probability. FLaG-MCMC therefore permits a flexible posterior representation based on a large reservoir without requiring the full kernel mixture to be evaluated at every MCMC transition. 

More precisely, each transition requires only a constant number of kernel and proposal-density evaluations together with
evaluation of the current data holder's likelihood. Denote by $C_L$ the cost of evaluating this likelihood. Typically, each kernel and proposal-density evaluation has cost $O(d)$, so the per-transition complexity is $O(d+C_L)$ and is independent of $B$. By contrast, a Metropolis random walk targeting the kernel-mixture posterior
directly incurs $O(Bd+C_L)$ computational cost per iteration because the complete mixture must be evaluated at each proposed state.

\subsection{The FLaG-MCMC Algorithm}
\label{sec:sequential-propagation}

Combining the kernel representation of
Section~\ref{sec:kernel-representation} with the graph-enabled MCMC transition of Section~\ref{sec:graph-mcmc} yields the complete FLaG-MCMC procedure. Starting from an initial reservoir $\mathcal R_0$, data holder $i$ applies the graph-enabled transition to target $\widehat\pi_i(\theta)\propto
\widetilde\pi_{i-1}(\theta)L_i(\theta)$. After burn-in, the $B$ retained posterior
draws of $\theta$ constitute the updated reservoir $\mathcal R_i$, which is returned to the central server for the next update. Repeating this procedure yields
\begin{align*}
\mathcal R_0
\longrightarrow
\mathcal R_1
\longrightarrow
\cdots
\longrightarrow
\mathcal R_N.
\end{align*}

\begin{figure}
\centering
\includegraphics[width=1.0\linewidth]{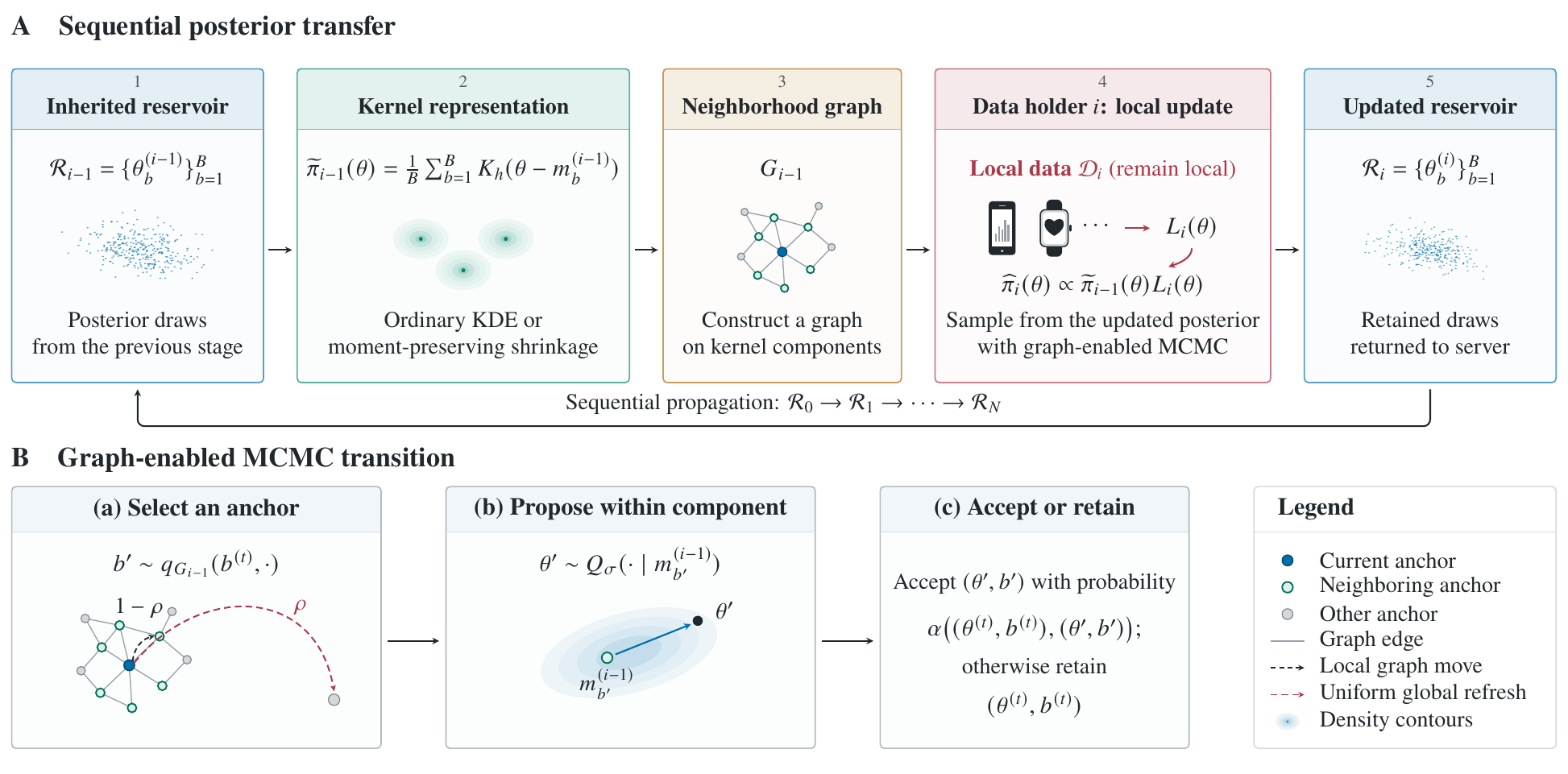}
\caption{Sequential posterior transfer and graph-enabled sampling in FLaG-MCMC.}
\label{fig:posterior_transfer_idea}
\end{figure}

Algorithm~\ref{alg:flag-mcmc} summarizes the complete procedure, and Figure~\ref{fig:posterior_transfer_idea} illustrates the sequential posterior transfer workflow and the graph-enabled MCMC transition performed at each stage. Models involving local latent variables are considered in the next subsection.

\begin{algorithm}
\caption{FLaG-MCMC for Sequential Posterior Transfer}
\label{alg:flag-mcmc}
\begin{algorithmic}[1]
\State Obtain an initial reservoir
$\mathcal R_0=\{\theta_b^{(0)}\}_{b=1}^{B}$ representing $\pi_0$
\For{data holder $i=1,\ldots,N$}
    \State Construct $\widetilde\pi_{i-1}$ from $\mathcal R_{i-1}$ using the
    chosen kernel representation
    \State Construct the neighborhood graph $G_{i-1}$ over the kernel
    components
    \State Initialize $(\theta^{(0)},b^{(0)})$
    \For{$t=0,\ldots,T-1$}
        \State Draw $b'\sim q_{G_{i-1}}(b^{(t)},\cdot)$
        \State Draw
        $\theta'\sim
        Q_\sigma(\cdot\mid m_{b'}^{(i-1)})$
        \State Accept $(\theta',b')$ with probability
        $\alpha\big((\theta^{(t)},b^{(t)}),(\theta',b')\big)$ in \eqref{eq:flag-mh}; otherwise retain
        $(\theta^{(t)},b^{(t)})$
    \EndFor
    \State After burn-in, retain $B$ posterior draws of $\theta$ to form
    $\mathcal R_i$ and return $\mathcal R_i$ to the server
\EndFor
\end{algorithmic}
\end{algorithm}

\subsection{Extension to Models with Local Latent Variables}
\label{sec:local-latent}

The preceding framework assumes that each data holder contributes through a likelihood term $L_i(\theta)$. In some models, however, data holder $i$ contains local latent variables $u_i$ that cannot be conveniently integrated out. The joint model then factorizes as
\begin{align*}
p(\theta,u_{1:N},\mathcal D_{1:N})
&=
\pi_0(\theta)
\prod_{i=1}^{N}
p(u_i\mid\theta)
p(\mathcal D_i\mid u_i,\theta).
\end{align*}
After incorporating the first $i$ data holders, the joint posterior for the global parameter $\theta$ and the local variables $u_i$ is
\begin{align*}
\pi_i(\theta,u_i)
:=p(\theta,u_i\mid\mathcal D_{1:i})
&\propto\pi_{i-1}(\theta)p(u_i\mid\theta)p(\mathcal D_i\mid u_i,\theta),
\end{align*}
where $\pi_{i-1}(\theta):=p(\theta\mid\mathcal D_{1:i-1})$ is the global posterior accumulated through the preceding $i-1$ data holders.

Under the federated constraint, $\pi_{i-1}(\theta)$ is available to data holder $i$ only through a finite posterior reservoir and is represented by the kernel approximation $\widetilde \pi_{i-1}(\theta)$. Replacing
$\pi_{i-1}(\theta)$ by this kernel representation gives the approximate
joint posterior
\begin{align*}
\widehat\pi_i(\theta,u_i)
&\propto
\widetilde \pi_{i-1}(\theta)
p(u_i\mid\theta)
p(\mathcal D_i\mid u_i,\theta).
\end{align*}
FLaG-MCMC samples from this distribution by retaining $u_i$ locally at data holder $i$ and updating it jointly with the global parameter through a Metropolis-within-Gibbs scheme.

The sampling procedure alternates between global and local updates. Conditional on
the current local state $u_i$, define
$\ell_i(\theta,u_i):=
p(u_i\mid\theta)p(\mathcal D_i\mid u_i,\theta)$. The graph-enabled update for $\theta$ proceeds as in Section~\ref{sec:graph-mcmc}, with $\ell_i(\theta,u_i)$ replacing
$L_i(\theta)$ in the Metropolis--Hastings ratio. The acceptance probability
is therefore
\begin{align*}
\alpha\big((\theta,b),(\theta',b')\big)
&=
\min\left\{
1,
\frac{
K_h(\theta'-m_{b'})
\ell_i(\theta',u_i)
q_G(b',b)
Q_\sigma(\theta\mid m_b)
}{
K_h(\theta-m_b)
\ell_i(\theta,u_i)
q_G(b,b')
Q_\sigma(\theta'\mid m_{b'})
}
\right\}.
\end{align*}

Conditional on the updated global parameter $\theta$, the local variables $u_i$ are
then updated using a Markov transition kernel that preserves the conditional distribution
\begin{align*}
\widehat\pi_i(u_i\mid\theta)
&\propto
p(u_i\mid\theta)
p(\mathcal D_i\mid u_i,\theta).
\end{align*}
Such an update can be implemented using, for example, Gibbs sampling or Metropolis--Hastings. Alternating the graph-enabled global update with the local update yields a
Metropolis-within-Gibbs sampler targeting
$\widehat\pi_i(\theta,u_i)$. After the MCMC computation at data holder $i$, only posterior draws of the shared parameter $\theta$ are returned to the central server; posterior draws of the local
variables $u_i$ and the observations $\mathcal D_i$ remain at data holder $i$.

\section{Theoretical Properties}
\label{sec:theory}

In this section, we establish theoretical properties of FLaG-MCMC. We first show that given the finite posterior reservoir, the graph-enabled Markov chain in Section~\ref{sec:graph-mcmc} has the desired stationary distribution $\widehat \pi_i$ as in~\eqref{poster_pi}.

\begin{theorem}[Stationarity]\label{T1}
For any stage $i\in [N]:=\{1,\ldots,N\}$, fix the posterior reservoir and its kernel representation. The graph-enabled Markov chain is reversible with respect to the augmented target $\widehat\Pi_i$ in~\eqref{pipostetheta}. Consequently,
the stationary distribution of $\theta$ is
$\widehat\pi_i$.
\end{theorem}

Next, we establish the consistency of sequential posterior transfer under appropriate regularity conditions, showing that $\widehat \pi_i$ converges to the pooled oracle posterior $\pi_i$ at each stage $i$. We state the result for the ordinary kernel representation; an analogous consistency result can be established for the moment-preserving shrinkage representation under suitable regularity conditions. For any two probability measures $P$ and $Q$, we denote by $d_{\mathrm{TV}}(P,Q)$ their total variation distance. We make the following assumptions.

\begin{assumption}\label{Assp0}
The kernel satisfies $\int_{\mathbb{R}^d} K(u)^2d u<\infty$. Moreover, there exist $M,\delta,\eta>0$ such that 
\begin{equation*}
    \sup_{\theta\in\mathbb{R}^d}\{L_i(\theta)\} \leq M\quad\text{ and }\quad \int_{\mathbb{R}^d}\pi_{i-1}(\theta)L_i(\theta)d\theta\geq \delta\quad\text{ for all }i\in [N],
\end{equation*}
\begin{equation}\label{domination}
    Q_{\sigma}(\theta\mid \theta_0)\geq \eta K_h(\theta-\theta_0)\quad\text{ for all }\theta,\theta_0\in \mathbb{R}^d.
\end{equation}
\end{assumption}
\begin{remark}
Condition~\eqref{domination} is satisfied with $\eta=1$ when we take the proposal kernel $Q_{\sigma}(\cdot\mid\theta_0)=K_h(\cdot-\theta_0)$ for every $\theta_0\in\mathbb{R}^d$.
\end{remark}

\begin{assumption}\label{assm}
The initial reservoir $\mathcal R_0=\{\theta_b^{(0)}\}_{b=1}^{B}$ consists of i.i.d. draws from the initial prior $\pi_0$.     
\end{assumption}

\begin{theorem}[Consistency of sequential targets]\label{T2}
Fix any $N<\infty$, and suppose that Assumptions~\ref{Assp0} and \ref{assm} hold. Additionally, assume that $h\to 0$ and $B h^d\to \infty$ as $B\to\infty$. Then for every $i\in [N]$, we have
\begin{align*}
     \mathbb{E}[d_{\mathrm{TV}}(\widehat\pi_i,\pi_i)]\to 0 \quad \text{ as }B\to \infty.
\end{align*}
\end{theorem}

Finally, we study how quickly the graph-enabled Markov chain approaches its stationary distribution. The total variation mixing time characterizes the number of iterations required for the Markov chain to be close to stationarity. Specifically, for any $\varepsilon\in(0,1)$, the total variation mixing time at
stage $i$ is
\begin{align*}
t_{\mathrm{mix},i}(\varepsilon)
:=
\min
\left\{
t\in\mathbb N:
\sup_{(\theta,b)\in\mathbb{R}^d\times [B]}
d_{\mathrm{TV}}
\left(
P_i^t((\theta,b),\cdot),
\widehat\Pi_i
\right)
\leq
\varepsilon
\right\},
\end{align*}
where $P_i$ denotes the transition kernel of the graph-enabled Markov chain, and $P_i^t((\theta,b),\cdot)$ denotes its distribution after $t$ iterations when initialized at $(\theta,b)$. The following result provides an explicit mixing time upper bound.  

\begin{theorem}[Mixing time bound]\label{T3}
\label{thm:flag-mixing} 
Assume that condition~\eqref{domination} holds. For any stage $i\in [N]$, suppose that $\sup_{\theta\in\mathbb{R}^d}
\{L_i(\theta)\}
<\infty$ and $\int_{\mathbb{R}^d}\widetilde\pi_{i-1}(\theta) L_i(\theta)d\theta>0$. Then, for every
$\varepsilon\in(0,1)$, 
\begin{align*}
t_{\mathrm{mix},i}(\varepsilon)
\leq
\left\lceil
\frac{\sup_{\theta\in\mathbb{R}^d}\{L_i(\theta)\}}{
\rho\eta\int_{\mathbb{R}^d}\widetilde\pi_{i-1}(\theta) L_i(\theta)d\theta
}
\log(\varepsilon^{-1})
\right\rceil.
\end{align*}
\end{theorem}
\begin{remark}
Under the conditions of Theorem~\ref{T2}, it can be shown that the mixing time bound in Theorem~\ref{T3} is bounded above by
\begin{equation}\label{ubbd}
\frac{\sup_{\theta\in\mathbb{R}^d}\{L_i(\theta)\}}{
\rho\eta\int_{\mathbb{R}^d}\pi_{i-1}(\theta) L_i(\theta)d\theta
}
\log(\varepsilon^{-1})+2
\end{equation}
with high probability. Thus, asymptotically, the mixing time bound remains bounded as the reservoir size $B$ increases.
\end{remark}

\section{Simulation Study: Federated Bayesian Meta-Analysis of Opioid Use Disorder Prevalence}
\label{sec:exp-oud}

In this section, we evaluate FLaG-MCMC in a federated Bayesian meta-analysis motivated by opioid use disorder (OUD) prevalence estimation. OUD has been prevalent in the United States for decades and continues to pose a major public health challenge \citep{blanco2019management,strang2020opioid}. Estimating the prevalence of untreated OUD is particularly difficult because relevant data are scarce, often distributed across health systems and jurisdictions, and may be subject to restrictions on direct sharing of individual-level records. In such settings, sharing posterior summaries or Monte Carlo draws provides a
practical means of transferring information from an existing Bayesian analysis without sharing the underlying individual-level data.

\citet{doogan2022opioid} developed a Bayesian framework for estimating OUD
prevalence in Ohio using linked health-record data, including a logistic regression model for the risk of fatal overdose among individuals with
untreated OUD. Building on this setting, our simulation study represents a subsequent analysis of untreated OUD prevalence in New York that incorporates information from the Ohio study without access to the underlying Ohio observations. The Ohio posterior is communicated through Monte Carlo draws and serves as informative
prior information for the New York analysis. 

We evaluate three aspects of FLaG-MCMC in this setting. We first examine how incorporating the transferred Ohio posterior changes inference in New York relative to an analysis using an uninformative prior. We then compare the resulting posterior with the pooled oracle, obtained by directly combining the Ohio and New York data and therefore unavailable under the federated constraint, to assess the approximation induced by finite-reservoir posterior transfer. Finally, we compare FLaG-MCMC with a Metropolis random walk targeting the same finite kernel-mixture posterior to assess the computational benefit of the graph-enabled transition.

\subsection{Experimental Setup}

For an individual with untreated OUD, let $Y_i\in\{0,1\}$ indicate fatal overdose and let $X_i\in\mathbb R^d$ denote a vector of individual characteristics. We consider the logistic regression model
\begin{align*}
Y_i\mid X_i,\beta
&\sim
\operatorname{Bernoulli}
\Big(
\operatorname{logit}^{-1}(X_i^\top\beta)
\Big),
\end{align*}
where $\operatorname{logit}^{-1}(t):=e^t/(1+e^t)$ and $\beta\in\mathbb R^d$ denotes the common regression parameter.

We generate independent Ohio and New York samples, each containing $1{,}500$ observations, with
\begin{align*}
X_i^{\mathrm{OH}}
\sim
N(\mathbf 1_d,I_d),\qquad
X_i^{\mathrm{NY}}
\sim
N(-\mathbf 1_d,I_d),
\end{align*}
where $\mathbf 1_d$ denotes the $d$-dimensional vector of ones. Conditional on the covariates, outcomes in both populations follow the logistic regression model above with the same parameter $\beta$. The two populations therefore share the same conditional model while exhibiting substantial covariate shift. We use the prior $\beta\sim N(0,I_d)$ and
consider dimensions $d\in\{2,6,10\}$.

The Ohio data are first used to obtain the posterior $\pi_{\mathrm{OH}}(\beta)$. We generate a reservoir $\mathcal R_{\mathrm{OH}}
=
\{\beta_b^{\mathrm{OH}}\}_{b=1}^{B}$
of $B=10{,}000$ posterior draws using \texttt{rjags}, with $40{,}000$ iterations and the first $30{,}000$ discarded as burn-in. The New York analysis receives $\mathcal R_{\mathrm{OH}}$ rather than the underlying Ohio observations. Because this experiment involves a single posterior transfer, we use the ordinary kernel representation introduced in Section~\ref{sec:ordinary-kernel}, with Gaussian kernel $K(u)=(2\pi)^{-d/2}\exp(-\|u\|^2/2)$ and bandwidth $h=0.04$. The resulting informative prior is
\begin{align*}
\widehat\pi_{\mathrm{OH}}(\beta)
&=
\frac{1}{B}
\sum_{b=1}^{B}
K_h\left(\beta-\beta_b^{\mathrm{OH}}\right),
\end{align*}
and combining this prior with the New York likelihood gives
\begin{align*}
\widehat\pi_{\mathrm{NY}}(\beta)
&\propto
\widehat\pi_{\mathrm{OH}}(\beta)
\prod_{i=1}^{1500}
\frac{
\exp\left(Y_i^{\mathrm{NY}}(X_i^{\mathrm{NY}})^\top\beta\right)
}{
1+\exp\left((X_i^{\mathrm{NY}})^\top\beta\right)
}.
\end{align*}

We compare two samplers targeting $\widehat\pi_{\mathrm{NY}}$. FLaG-MCMC
uses the graph-enabled transition described in Section~\ref{sec:graph-mcmc}, whereas the comparator is a Metropolis random walk that evaluates the full $B$-component kernel mixture at every proposal and uses a Gaussian proposal with covariance $0.02^2 I_d$. For FLaG-MCMC, we set $k=\lceil\sqrt{B}\rceil$ and construct a symmetrized $k$-nearest-neighbor graph: an edge $\{b,b'\}$ is included if $\beta_b^{\mathrm{OH}}$ is among the $k$ nearest neighbors of $\beta_{b'}^{\mathrm{OH}}$, or vice versa. Additionally, we take $\rho=0.5$ and use a Gaussian proposal kernel $Q_\sigma$ with covariance $0.04^2 I_d$. Each sampler is run for $10{,}000$ iterations, with the first $5{,}000$ discarded as burn-in.

As an inferential benchmark, we obtain the pooled oracle posterior by directly combining the Ohio and New York data under the same Bayesian model and sampling from the resulting posterior using \texttt{rjags}. This posterior is
unavailable under the federated constraint and allows us to assess the approximation induced by finite-reservoir posterior transfer. 

\subsection{Posterior Transfer and Accuracy}

\begin{figure}
\centering
\includegraphics[width=0.88\linewidth]{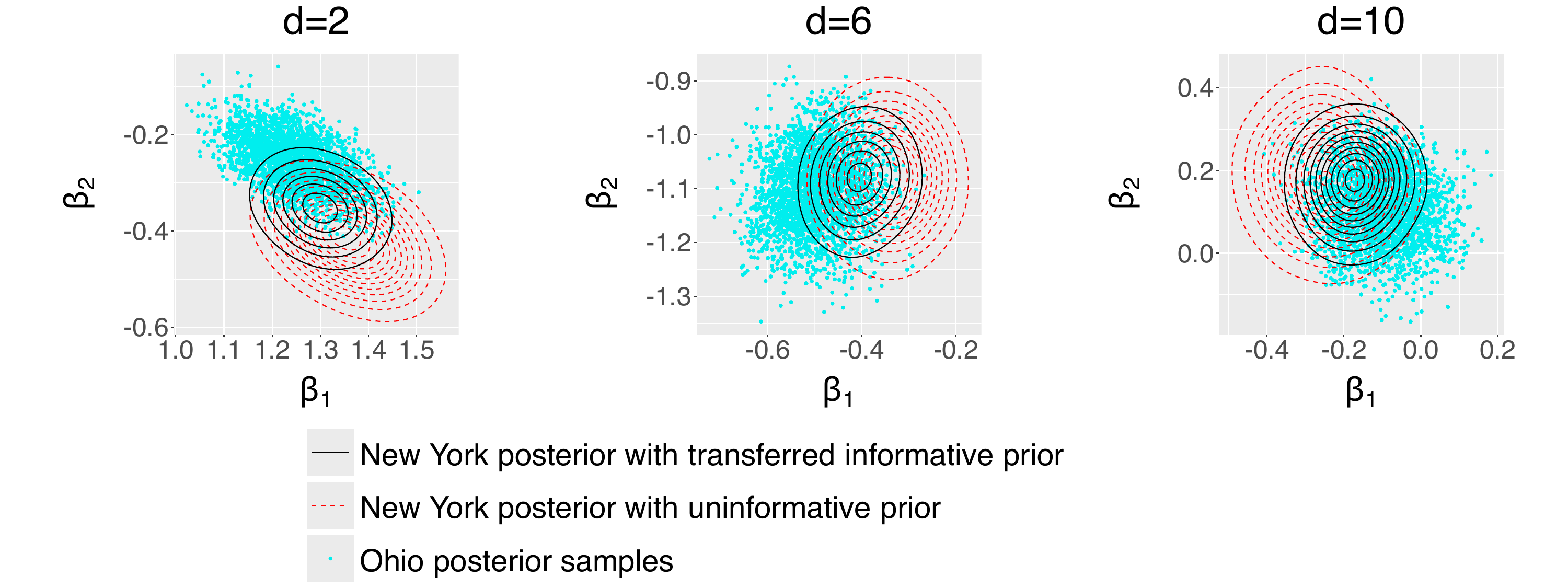}
\caption{Federated posterior transfer from Ohio to New York for $d=2,6,10$. Cyan points show samples from the Ohio posterior, solid black contours
represent the New York posterior obtained by FLaG-MCMC using the transferred
informative prior, and dashed red contours represent the New York posterior
under the uninformative prior.}
\label{fig:oud-contour}
\end{figure}

We first examine how transferring the Ohio posterior changes posterior inference in
New York and how closely the resulting finite-reservoir posterior approximates the pooled oracle. Figure~\ref{fig:oud-contour} illustrates the effect of posterior transfer for the first two coordinates of $\beta$. The New York posterior obtained by FLaG-MCMC using the transferred informative
prior lies between the Ohio posterior and the New York posterior under the uninformative prior, confirming that information from the Ohio study is successfully integrated into the New York analysis.

Table~\ref{tab:oud-performance} reports the $2$-Wasserstein distance \citep[Chapter 6]{villani2009optimal} between each method's estimated posterior and the pooled oracle. Both methods achieve small $2$-Wasserstein distances, indicating close approximation to the pooled oracle, with negligible differences between the two methods. 

\begin{table}
\caption{Posterior accuracy and computation time for the federated OUD
meta-analysis. The $2$-Wasserstein distance is computed relative to the
pooled oracle. Computation time is reported in seconds per MCMC iteration
for $B=10{,}000$. Entries are averaged over three replications.}
\label{tab:oud-performance}
\centering
\begin{tabular}{@{}lcccccc@{}}
\hline
& \multicolumn{3}{c}{$W_2$ distance}
& \multicolumn{3}{c}{Time per iteration} \\
\cline{2-4}\cline{5-7}
Method
& $d=2$ & $d=6$ & $d=10$
& $d=2$ & $d=6$ & $d=10$ \\
\hline
FLaG-MCMC
& 0.020 & 0.060 & 0.16
& 0.017 & 0.018 & 0.018 \\
Metropolis random walk
& 0.022 & 0.057 & 0.15
& 0.11 & 0.11 & 0.11 \\
\hline
\end{tabular}
\end{table}

\subsection{Computational Performance}

We next examine the computational efficiency of FLaG-MCMC and its scaling
with the posterior reservoir size $B$. Table~\ref{tab:oud-performance} shows
that, at $B=10{,}000$, FLaG-MCMC is approximately six times faster per iteration than the Metropolis random walk, with a similar computational advantage across $d=2,6,10$.

\begin{figure}[t]
\centering
\includegraphics[width=0.5\linewidth]{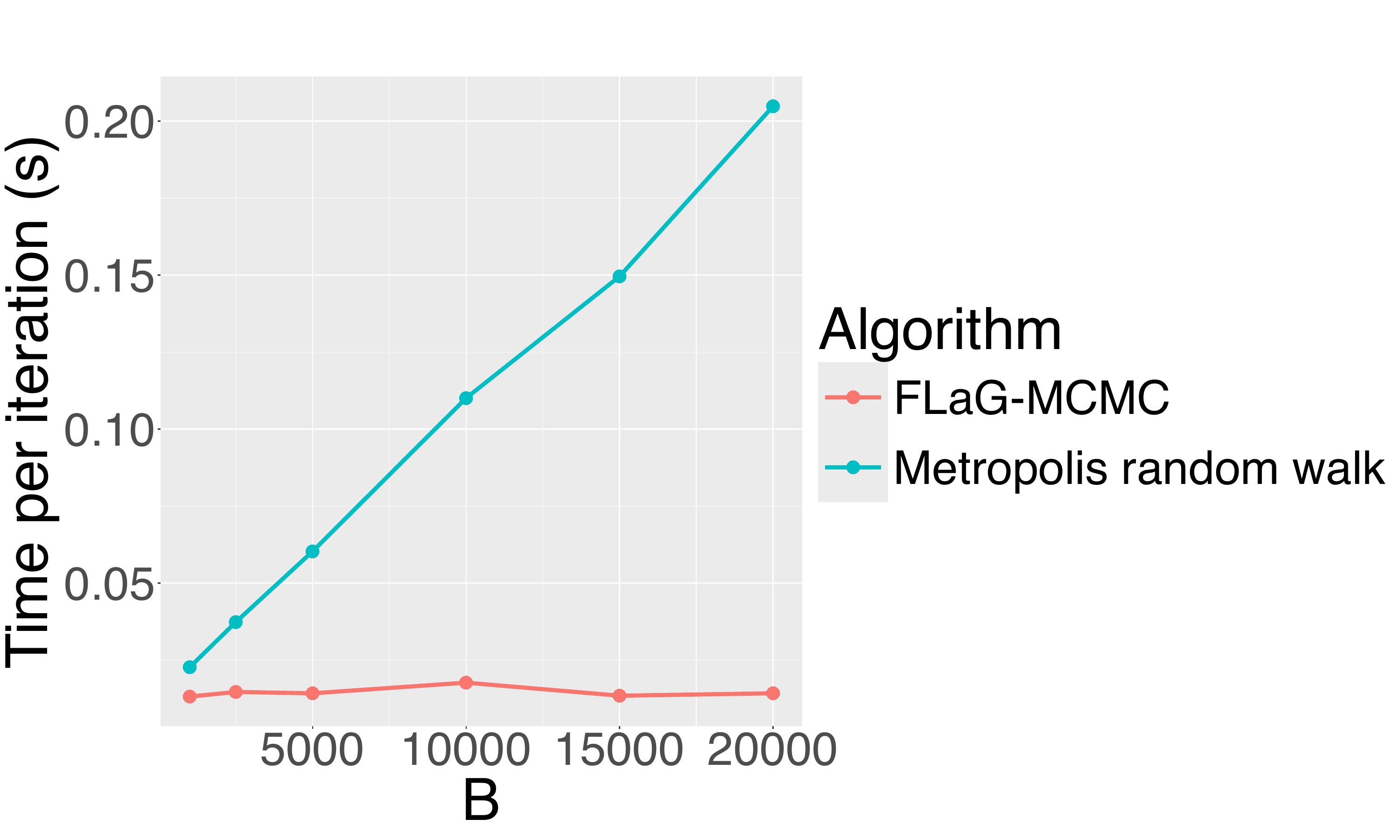}
\caption{Computation time per iteration as a function of the
posterior reservoir size $B$ for $d=6$, averaged over three replications.}
\label{fig:oud-runtime}
\end{figure}

To examine the dependence of the computational cost on $B$, we repeat the $d=6$ experiment with $B\in
\{1000,2500,5000,10000,15000,20000\}$,
using $k=\lceil\sqrt{B}\rceil$ and fixing the other tuning parameters as before. Figure~\ref{fig:oud-runtime} shows that the per-iteration cost of the Metropolis random walk increases approximately linearly with $B$, whereas that of FLaG-MCMC remains nearly constant. Across all values of $B$ considered, FLaG-MCMC has substantially lower computational cost than the Metropolis random walk. The observed dependence on $B$ agrees with the complexity analysis in Section~\ref{sec:graph-mcmc}: increasing the posterior reservoir size does not impose a corresponding increase in the cost of each FLaG-MCMC transition.

The computational gains of FLaG-MCMC are not achieved at the expense of MCMC performance. Both samplers attain multivariate potential scale reduction factors (MPSRFs) \citep{brooks1998general} close to one, indicating satisfactory mixing, while FLaG-MCMC yields approximately $2$--$3$ times larger effective sample sizes than the Metropolis random walk. Taken together, these results show that FLaG-MCMC combines substantially lower per-iteration computational cost with satisfactory mixing and higher statistical efficiency. Detailed convergence diagnostics, autocorrelation plots, and effective sample sizes are reported in Appendix~\ref{Appendix_B}.

\section{Application: Federated Bayesian Analysis of Depression Status Using Mobile and Wearable Data}
\label{sec:globem}

In this section, we apply FLaG-MCMC to participant-level data from GLOBEM,
a multi-year passive-sensing study combining measurements from smartphone and wearable devices with survey assessments. We propagate the posterior sequentially across $40$ participants, each contributing repeated depression assessments and behavioral summaries derived from passive sensing. This setting provides a substantially longer and more heterogeneous evaluation of sequential posterior transfer. FLaG-MCMC closely tracks the evolving sequence of pooled Bayesian posteriors throughout the $40$ participant-level updates, preserving posterior uncertainty as heterogeneous participant information is incorporated sequentially. Throughout the procedure, participant-level observations remain local, and only the global posterior representation is propagated between stages.

\subsection{Scientific Motivation and Data}

Smartphones and wearable devices provide increasingly important sources
of longitudinal behavioral information in health research. Passive sensing
can characterize mobility, daily routines, physical activity, sleep, and
device-use patterns in everyday settings
\citep{lane2010survey,wang2014studentlife}. A substantial literature has
linked such behavioral signals to mental health outcomes, including
depression \citep{saeb2015mobile,canzian2015trajectories}. At the same time, these measurements are inherently participant-level and potentially sensitive, making data-local analysis particularly relevant in mobile health applications.

We analyze GLOBEM \citep{xu2022globem}, a multi-year passive-sensing dataset developed to study human behavior and well-being across participants and study years. Data were collected for approximately ten weeks in each of four study years, 2018--2021. Smartphones recorded behavioral signals related to mobility and phone use, wearable devices recorded physical activity and sleep, and survey instruments provided
repeated measures of well-being and depression. The study exhibits substantial heterogeneity across participants and study years, including pronounced behavioral distribution shifts between the pre- and post-COVID periods.

Depression detection is a central modeling task considered in GLOBEM, where passive-sensing measurements are used to characterize behavioral patterns associated with depression status. We consider this problem from a Bayesian perspective. Rather than focusing solely on classification, our
analysis quantifies uncertainty in associations between depression status
and interpretable behavioral summaries derived from smartphone and wearable
measurements. This posterior uncertainty is itself an inferential object of interest and can evolve as heterogeneous participant-level information is incorporated sequentially.

For reproducibility, we use the publicly available GLOBEM sample containing $40$ participants, with ten participants from each study year. Each participant contributes three binary depression assessments, yielding $120$ assessment records in total. For assessment $j$ of participant $i$, behavioral features are constructed using sensor measurements from the strictly preceding 14-day window. Participants are processed in study-year order, beginning with the ten 2018 participants and proceeding through 2019, 2020, and 2021.

\subsection{Bayesian Hierarchical Model}

We formulate a Bayesian hierarchical logistic regression model motivated by
the sensor-based depression analysis of \citet{saeb2015mobile}. The model
uses interpretable smartphone-derived measures of mobility, location regularity, and patterns of phone use. GLOBEM \citep{xu2022globem} additionally provides wearable measurements of physical activity and sleep, which we incorporate together with study-year indicators. Repeated
assessments within participants are accommodated through a participant-specific random intercept.

Let $i=1,\ldots,40$ index participants and $j=1,2,3$ index repeated depression assessments. Let $Y_{ij}\in\{0,1\}$ denote the binary depression indicator supplied in the GLOBEM sample. For each assessment, treating $2018$ as the reference year, we define
\begin{align*}
x_{ij}
=
\left(
1,
w_{ij,1},\ldots,w_{ij,13},
I\{\mathrm{year}=2019\},
I\{\mathrm{year}=2020\},
I\{\mathrm{year}=2021\}
\right)^\top.
\end{align*}
The thirteen behavioral summaries $w_{ij,1},\ldots,w_{ij,13}$ are \emph{log location variance},
\emph{number of significant places}, \emph{location entropy},
\emph{normalized location entropy}, \emph{home time},
\emph{circadian routine}, \emph{stationary ratio},
\emph{distance travelled}, \emph{unlock duration},
\emph{unlock count}, \emph{steps}, \emph{sleep duration},
and \emph{sleep efficiency}. These summaries span the principal sensing domains represented in the application: mobility and location regularity, phone engagement, physical activity, and sleep. Detailed definitions are provided in Appendix~\ref{Appendix_C}.

We use the hierarchical logistic regression model
\begin{align*}
Y_{ij}\mid b_i,\beta
&\sim
\operatorname{Bernoulli}(p_{ij}),
\\
\operatorname{logit}(p_{ij})
&=
x_{ij}^\top\beta+b_i,
\\
b_i\mid\tau
&\sim
N(0,\tau^2),
\end{align*}
where $b_i$ is a participant-specific random intercept that accounts for
within-participant dependence among the three repeated assessments, and $\operatorname{logit}(x):=\log\big(\frac{x}{1-x}\big)$ for $x\in (0,1)$. The
fixed-effect vector is $\beta
=
(\beta_0,\beta_1,\ldots,\beta_{16})^\top$, with one intercept, thirteen behavioral coefficients, and three study-year
indicators. We place independent Gaussian priors
\begin{align*}
    \beta_0 \sim N(0,1.5^2),\quad \beta_r \sim N(0,0.75^2)  \text{  for }  r=1,\ldots,16, \quad \eta=\log\tau\sim N(\log(0.75),0.5^2).
\end{align*}
The globally shared parameter is therefore $\theta
=(\beta^\top,\eta)^\top\in\mathbb R^{18}$.

The participant-specific random intercept $b_i$ is a local latent variable
and is not included in the posterior reservoir propagated across
participants. For convenience, define
\begin{align*}
p_{ij}(b;\beta)
=
\operatorname{logit}^{-1}
\Big(
x_{ij}^\top\beta+b
\Big).
\end{align*}
We marginalize the random intercept from the participant likelihood,
\begin{align*}
L_i(\beta,\tau)
=
\int
\left[
\prod_{j=1}^{3}
p_{ij}(b;\beta)^{Y_{ij}}
\left(
1-p_{ij}(b;\beta)
\right)^{1-Y_{ij}}
\right]
\phi(b;0,\tau^2)\,db.
\end{align*}
We evaluate this one-dimensional integral using 25-point Gauss--Hermite quadrature. Let $(u_g,w_g)$, $g=1,\ldots,25$, denote the quadrature nodes and weights. Then
\begin{align*}
L_i(\theta)
\approx
\frac{1}{\sqrt{\pi}}
\sum_{g=1}^{25}
w_g
\prod_{j=1}^{3}
p_{ij}\left(\sqrt{2}\tau u_g;\beta\right)^{Y_{ij}}
\left(  1-p_{ij}\left(\sqrt{2}\tau u_g;\beta\right)
\right)^{1-Y_{ij}}.
\end{align*}
Thus each participant contributes a likelihood for the common global parameter $\theta$, while the participant-specific latent effect is integrated out before posterior information is propagated.

At sequential stage $s$, the pooled-oracle posterior is $\pi_s^{\mathrm{oracle}}(\theta)
\propto
p(\theta)
\prod_{i=1}^{s}
L_i(\theta)$, where $p(\theta)$ denotes the prior density and $L_i(\theta)$ the likelihood contribution from participant $i$. This is the centralized Bayesian posterior used as the reference throughout
the application and is sampled using random-walk Metropolis--Hastings. At stage 1, FLaG-MCMC is initialized by sampling directly from $\pi_1^{\mathrm{oracle}}$. For $s\geq2$, FLaG-MCMC updates the finite posterior representation inherited from the preceding stage. Conditional on the
stage-$(s-1)$ reservoir, let
$\widetilde\pi_{s-1}^{\mathrm{FLaG}}$ denote its shrinkage kernel
representation. The stage-$s$ FLaG-MCMC target is $\widehat\pi_s^{\mathrm{FLaG}}(\theta)
\propto
\widetilde\pi_{s-1}^{\mathrm{FLaG}}(\theta) L_s(\theta)$. We assess the fidelity of the FLaG-MCMC posterior by comparing it with $\pi_s^{\mathrm{oracle}}$ at every sequential stage.

\subsection{Sequential Implementation and Results}

We initialize the sequential analysis with the first participant and subsequently propagate the global posterior across all $40$ participants. At each stage, FLaG-MCMC receives the posterior reservoir from the preceding stage, incorporates the current participant likelihood, and returns an updated reservoir of size $B=10^6$. Because the posterior is repeatedly represented and transferred across many participants, we use the moment-preserving shrinkage kernel representation introduced in Section~\ref{sec:shrinkage}. Further implementation details are provided in Appendix~\ref{Appendix_C}.

For FLaG-MCMC, we run four independent chains at every stage. Each chain discards $10^4$ burn-in iterations and retains $2.5\times10^5$ post-burn-in draws, yielding the $10^6$-draw posterior reservoir used for the next participant. For the pooled oracle, we likewise run four independent chains, each with $10^4$ burn-in iterations and $2.5\times10^5$ retained draws.

FLaG-MCMC required approximately $7.6$ minutes of wall-clock time for all $40$ sequential participant updates on a 2021 MacBook Pro with an Apple M1 Pro chip and
16 GB of memory. We assess convergence using MPSRF, following the same diagnostic used in Section~\ref{sec:exp-oud}. Across all $40$ stages, the MPSRF remains below
$1.005$ for both FLaG-MCMC and the pooled oracle, indicating strong convergence of the four chains throughout the sequential analysis.

\begin{figure}
\centering
\includegraphics[width=0.90\linewidth]{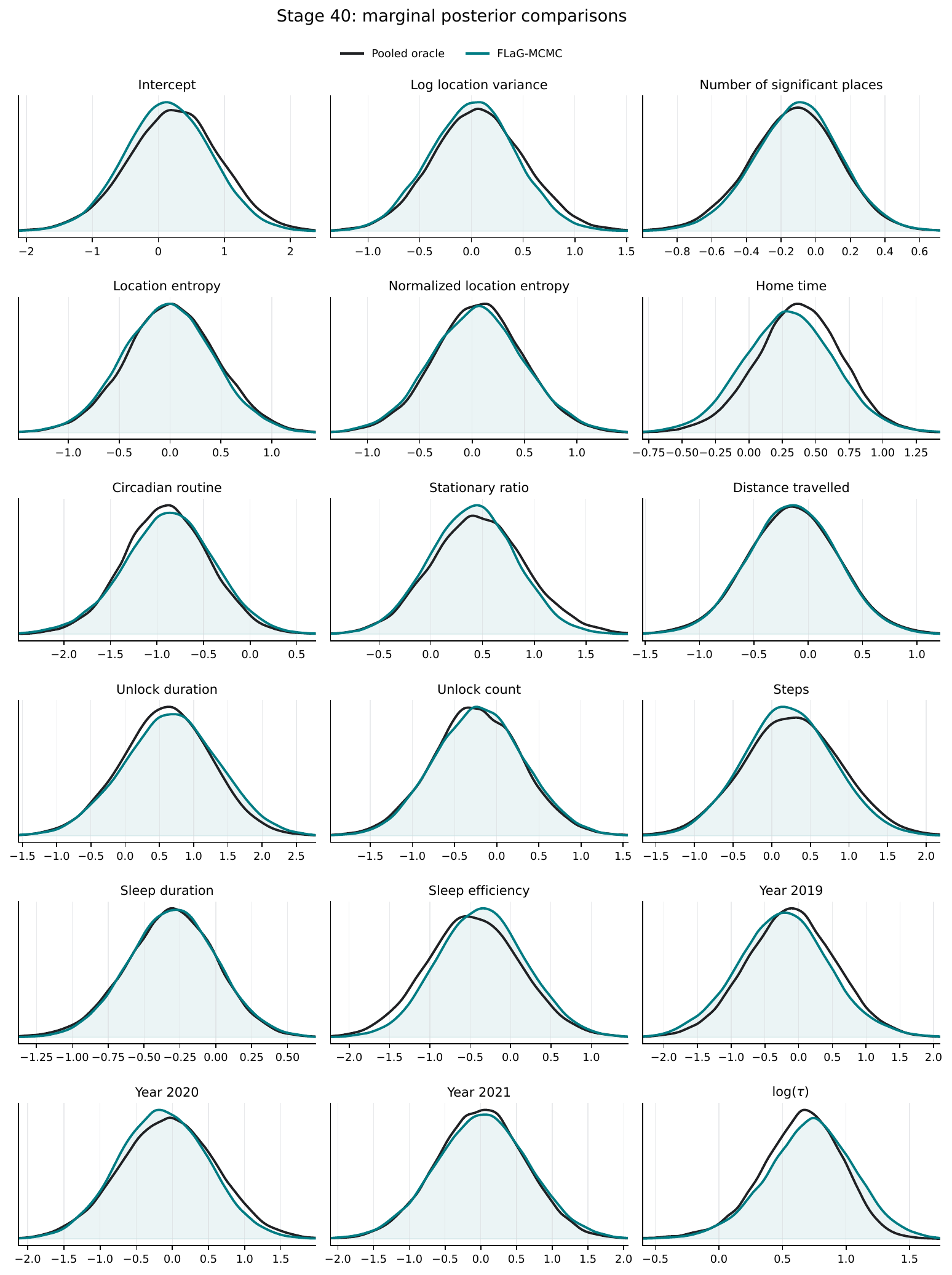}
\caption{Marginal posterior comparisons after all $40$ participants (2018--2021). Black curves denote the pooled oracle and teal curves denote
FLaG-MCMC. The close agreement across parameters indicates that the sequentially propagated posterior closely reproduces the pooled Bayesian posterior after all $40$ participant updates.}
\label{fig:stage40}
\end{figure}

We first examine the posterior after all $40$ participants and four study years have been incorporated. Figure~\ref{fig:stage40} compares the marginal posterior distributions obtained by FLaG-MCMC with those from the pooled oracle. FLaG-MCMC closely reproduces the location, concentration, and overall shape of the pooled posterior across the global parameters. This agreement is notable because the FLaG-MCMC posterior has undergone $40$ successive participant-level Bayesian updates and repeated nonparametric posterior representation. Complete marginal comparisons at intermediate stages 10,
20, and 30 are provided in Appendix~\ref{Appendix_C} and show similarly
close agreement throughout the sequential analysis.

Beyond the above marginal comparisons, we quantify posterior discrepancy between FLaG-MCMC and the pooled oracle throughout the sequential analysis, considering both marginal and multivariate posterior agreement. For each parameter $j$ at stage $s$, we compute the one-dimensional 1-Wasserstein distance between the FLaG-MCMC and pooled-oracle marginal posteriors and
standardize it by the corresponding pooled-oracle posterior standard deviation,
\begin{align*}
\widetilde W_{1,sj}
=
\frac{
W_1\left(
\widehat\pi_{s,j}^{\mathrm{FLaG}},
\widehat\pi_{s,j}^{\mathrm{oracle}}
\right)
}{
\widehat\sigma_{s,j}^{\mathrm{oracle}}
}.
\end{align*}
We summarize marginal discrepancy by averaging $\widetilde W_{1,sj}$ over
the 18 global parameters. This standardization places parameters with
different posterior scales on a common scale.

For the multivariate comparison, we consider the sliced 1-Wasserstein
distance \citep{rabin2011wasserstein,chewi2025statistical}. For probability distributions $P$ and $Q$ on $\mathbb R^d$, define
\begin{align*}
\operatorname{SW}_1(P,Q)
=
\int_{\mathbb S^{d-1}}
W_1(P_u,Q_u)
\,d\sigma(u),
\end{align*}
where $P_u$ and $Q_u$ denote the one-dimensional distributions obtained by
projecting $P$ and $Q$ onto direction $u$, and $\sigma$ denotes the uniform probability measure on the unit sphere $\mathbb S^{d-1}$. To place the multivariate comparison on a common scale, we whiten both posterior samples using the pooled-oracle posterior at each stage. Let
$\widehat\mu_s^{\mathrm{oracle}}$ and
$\widehat\Sigma_s^{\mathrm{oracle}}=C_sC_s^\top$ denote the empirical mean and covariance of the pooled-oracle draws at stage $s$. We define
\begin{align*}
z_i^{\mathrm{oracle}}
=
C_s^{-1}
\left(
\theta_i^{\mathrm{oracle}}
-
\widehat\mu_s^{\mathrm{oracle}}
\right),\qquad
z_i^{\mathrm{FLaG}}
=
C_s^{-1}
\left(
\theta_i^{\mathrm{FLaG}}
-
\widehat\mu_s^{\mathrm{oracle}}
\right).
\end{align*}
We estimate the sliced 1-Wasserstein distance between the resulting whitened
distributions using $M$ projection directions $u_1,\ldots,u_M$ drawn
uniformly from $\mathbb S^{d-1}$,
\begin{align*}
\widehat{\operatorname{SW}}_{1,s}
=
\frac{1}{M}
\sum_{m=1}^{M}
W_1\left(
u_m^\top z^{\mathrm{FLaG}},
u_m^\top z^{\mathrm{oracle}}
\right).
\end{align*}
We use $M=500$ projection directions and randomly subsample $30{,}000$ draws from each whitened posterior sample for this calculation. We refer to this oracle-whitened quantity as the standardized
sliced 1-Wasserstein distance. 

\begin{figure}
\centering
\includegraphics[width=0.75\linewidth]{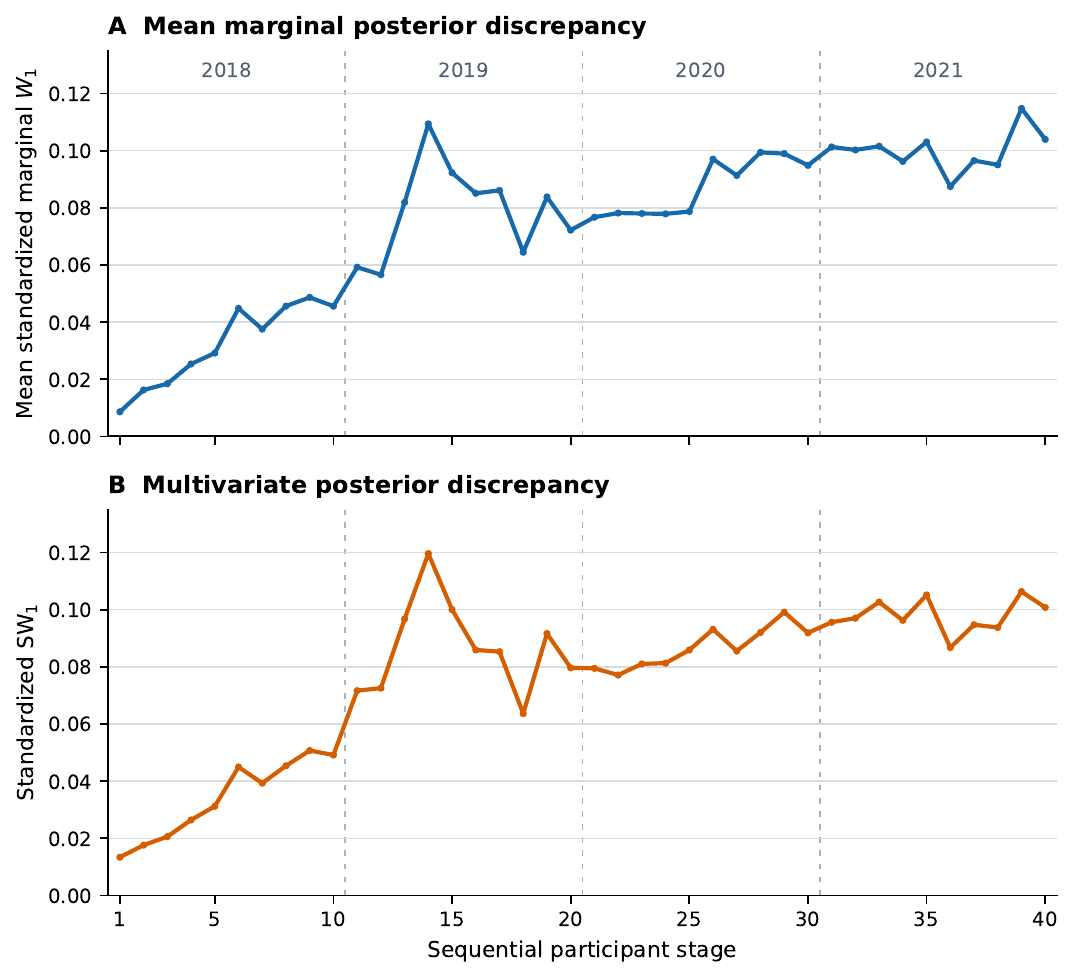}
\caption{Posterior fidelity across sequential participant updates.
Panel A shows the mean standardized marginal 1-Wasserstein distance between
FLaG-MCMC and the pooled oracle. Panel B shows the standardized sliced
1-Wasserstein distance, computed after common whitening by the pooled-oracle
posterior. Dashed vertical lines indicate study-year boundaries.}
\label{fig:globem_fidelity}
\end{figure}

Figure~\ref{fig:globem_fidelity} reports both discrepancies across the
sequential participant updates. Across stages 2--40, the mean standardized
marginal 1-Wasserstein distance averages $0.0763$, with a maximum stagewise value of $0.1148$. The standardized sliced 1-Wasserstein distance has a corresponding
average of $0.0779$ and a maximum of $0.1197$. Both marginal and multivariate posterior discrepancies therefore remain small throughout the sequential analysis. After increasing during the earlier stages, both measures fluctuate within a relatively stable range through the remainder of the participant sequence. Together with the final-stage marginal comparison shown in Figure~\ref{fig:stage40}, these results show that FLaG-MCMC remains close to the pooled Bayesian posterior throughout all $40$ participant updates.

\begin{figure}
\centering
\includegraphics[width=0.85\linewidth]{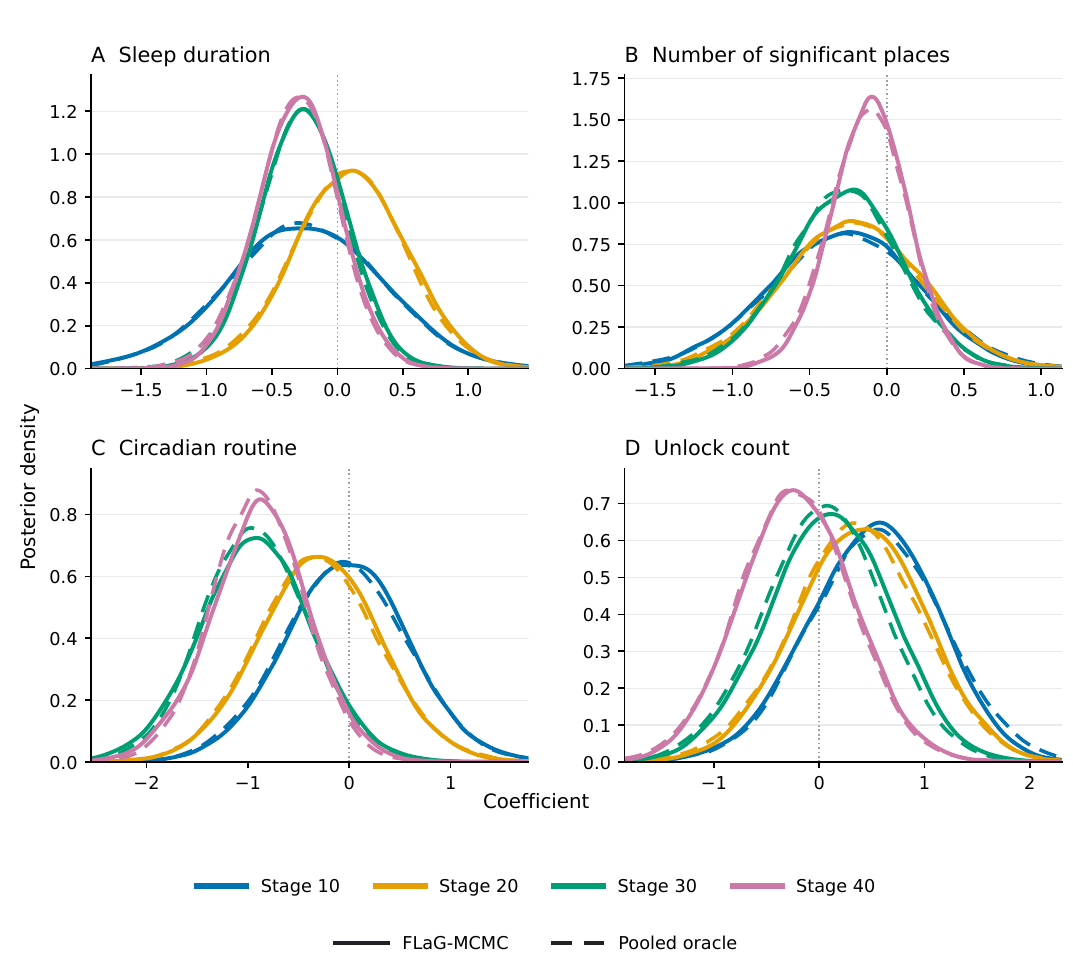}
\caption{Evolution of marginal posterior distributions for four behavioral coefficients at stages 10, 20, 30, and 40. The coefficients correspond to \emph{sleep duration}, \emph{number of significant places}, \emph{circadian routine}, and \emph{unlock count}, representing sleep, mobility breadth, daily mobility regularity, and phone usage, respectively. Solid curves represent FLaG-MCMC and dashed curves represent the pooled oracle. Across stages, the FLaG-MCMC marginals closely track their pooled-oracle counterparts while the posterior distributions evolve as participant information accumulates.}
\label{fig:globem_marginal_evolution}
\end{figure}

The evolving posterior is also scientifically informative.
Figure~\ref{fig:globem_marginal_evolution} follows the marginal posteriors
for the \emph{sleep duration}, \emph{number of significant places}, \emph{circadian routine}, and \emph{unlock count} coefficients at stages 10, 20, 30, and 40. The trajectories reveal substantial evolution in both posterior location and uncertainty as participant information accumulates. For \emph{sleep duration}, the posterior shifts rightward between stages
10 and 20 and then moves leftward and becomes more concentrated at stages
30 and 40. The posterior for \emph{number of significant places} becomes more concentrated at stage 40. Both the \emph{circadian routine} and \emph{unlock count} coefficients shift across stages.
These patterns illustrate that the posterior being propagated can change
substantially as heterogeneous participants and study years are incorporated. FLaG-MCMC closely reproduces these changes at each displayed stage while operating under the federated constraint, tracking the corresponding pooled-oracle marginals as both posterior location and uncertainty evolve.

Taken together, the GLOBEM analysis demonstrates both posterior fidelity and meaningful posterior evolution across the sequential analysis. FLaG-MCMC closely tracks the pooled Bayesian posterior throughout $40$ participant-level updates, with strong Markov chain convergence and
small marginal and multivariate posterior discrepancies. At the same time,
Figure~\ref{fig:globem_marginal_evolution} shows that the posterior being
transferred can change substantially in both location and uncertainty as
heterogeneous participant information is incorporated. These results demonstrate that participant-level federated Bayesian analysis using FLaG-MCMC can retain evolving posterior uncertainty while closely tracking the corresponding pooled Bayesian analysis.

\section{Discussion and Conclusions}
\label{sec:discussion}

We have introduced FLaG-MCMC, a framework for federated Bayesian modeling through sequential posterior transfer across distributed data holders. Rather than reducing accumulated information to a point estimate or parametric summary, the framework propagates a nonparametric reservoir of posterior draws, retaining uncertainty quantification and joint posterior structure as new data holders contribute information. The graph-enabled sampler makes this sample-based posterior representation computationally tractable by exploiting the geometry of the reservoir, while the moment-preserving kernel representation supports repeated transfer across long sequences of data holders. Together, these components enable efficient accumulation of Bayesian evidence in federated settings. 

The two health science applications illustrate federated Bayesian evidence synthesis at different levels of data decentralization. In the GLOBEM analysis, posterior information is accumulated across individual participants, while participants' underlying mobile and wearable observations remain local. FLaG-MCMC closely reconstructs the corresponding pooled Bayesian posterior throughout the participant sequence while tracking changes in the location and uncertainty of the posterior distributions for the behavioral associations. The OUD study considers evidence synthesis across studies, transferring the posterior from an Ohio study to a subsequent New York analysis under covariate shift. The close agreement of FLaG-MCMC with the pooled posterior, together with favorable computational scaling relative to the Metropolis random walk, demonstrates reliable and efficient posterior transfer in the meta-analysis setting. These applications illustrate how the same framework can support federated Bayesian analysis when evidence is distributed across individual participants, studies, or populations.

Several extensions could further broaden the scope of FLaG-MCMC. For high-dimensional global parameters with complex posterior geometry, a lower-dimensional latent embedding can provide a more effective representation of the posterior structure for graph construction and proposal moves. For example, an encoder--decoder architecture could learn a representation adapted to the posterior geometry, with graph construction and proposal moves performed in the latent space and the decoder mapping proposed states back to the original parameter space for likelihood evaluation. When scientifically interpretable low-dimensional structure is available, the representation could instead be constructed from a structured parameterization or a prespecified transformation of the global parameter.

More broadly, flexible sample-based posterior representation and graph-enabled sampling provide a general framework
for Bayesian evidence synthesis across a wide range of scientific applications with decentralized data. Potential applications include multi-institution biomedical studies, longitudinal and multi-site health research, and transfer learning across related populations. The framework also opens avenues for future research on sequential posterior transfer in a broad class of Bayesian models with complex global and local structure, including mixed membership models \citep{airoldi2014handbook} and Bayesian nonparametric models \citep{hjort2010bayesian}. 

\bibliographystyle{plainnat}
\bibliography{Graph-MCMC}

\appendix
\newpage

\section{Proofs of Theoretical Results}\label{Appendix_A}

In this section, we present the proofs of Theorems~3.1--3.3.  

\subsection{Proof of Theorem~3.1}

Recall the notation in Section~2.3 of the main paper. Given the current state $(\theta,b)\in\mathbb{R}^d\times[B]$, the proposal density for $(\theta',b')\in\mathbb{R}^d\times[B]$ is \begin{equation*}
    \mathcal{Q}\big((\theta,b),(\theta',b')\big)=q_G(b,b')Q_{\sigma}(\theta'\mid m_{b'}).
\end{equation*}  
The corresponding Metropolis--Hastings acceptance probability with respect to $\widehat\Pi_i$ is
\begin{equation*}
    \min\left\{1,\frac{\widehat\Pi_i(\theta',b')\mathcal{Q}\big((\theta',b'),(\theta,b)\big)}{\widehat\Pi_i(\theta,b)\mathcal{Q}\big((\theta,b),(\theta',b')\big)}\right\}=\min\left\{
1,\frac{
K_h(\theta'-m_{b'})
L_i(\theta')
q_G(b',b)
Q_\sigma(\theta\mid m_b)
}{
K_h(\theta-m_b)
L_i(\theta)
q_G(b,b')
Q_\sigma(\theta'\mid m_{b'})
}
\right\},
\end{equation*}
which is the acceptance probability in Equation~(3) of the main paper. Hence the graph-enabled Markov chain is reversible with respect to $\widehat\Pi_i$. Moreover, 
\begin{align*}
\sum_{b=1}^{B}
\widehat \Pi_i(\theta,b)
&\propto
\frac{1}{B}
\sum_{b=1}^{B}
K_h(\theta-m_b)
L_i(\theta)
\propto
\widehat\pi_i(\theta).
\end{align*}
Therefore, the stationary distribution of $\theta$ is $\widehat \pi_i$.

\subsection{Proof of Theorem~3.3}

We denote by $P_i$ the transition kernel of the graph-enabled Markov chain at stage $i$. For any $b,b'\in[B]$, $\theta\in\mathbb{R}^d$, and any Borel measurable set $A\subseteq\mathbb{R}^d$, by Equation~(3) in the main paper, we have 
\begin{align*}
   & P_i\big((\theta,b),\{(\theta',b'):\theta'\in A\}\big)\geq \int_A q_G(b,b') Q_{\sigma}(\theta'\mid m_{b'})\alpha\big((\theta,b),(\theta',b')\big)d\theta'\nonumber\\
   =\,& \int_A \min\left\{q_G(b,b')Q_{\sigma}(\theta'\mid m_{b'}),\frac{
K_h(\theta'-m_{b'})
L_i(\theta')
q_G(b',b)
Q_\sigma(\theta\mid m_b)
}{
K_h(\theta-m_b)
L_i(\theta)
}\right\} d\theta'\nonumber\\
\geq\,& \frac{\rho\eta}{B}\int_A K_h(\theta'-m_{b'}) \min\left\{1,\frac{L_i(\theta')}{L_i(\theta)}\right\}d\theta'\geq \frac{\rho\eta}{B\sup_{\gamma\in\mathbb{R}^d}\{L_i(\gamma)\}}\int_A K_h(\theta'-m_{b'})L_i(\theta') d\theta'
\end{align*}
where the third line uses $q_G(b,b')\geq \rho/B$, $q_G(b',b)\geq \rho/B$, and condition~(4) in the main paper. Moreover,
\begin{align*}
    \widehat\Pi_i\big(\{(\theta',b'):\theta'\in A\}\big)=\frac{\int_A K_h(\theta'-m_{b'})L_i(\theta')d\theta'}{\sum_{c=1}^B\int_{\mathbb{R}^d}K_h(\gamma-m_c)L_i(\gamma)d\gamma}=\frac{\frac{1}{B}\int_A K_h(\theta'-m_{b'})L_i(\theta')d\theta'}{\int_{\mathbb{R}^d}\widetilde\pi_{i-1}(\gamma) L_i(\gamma)d\gamma}.
\end{align*}
Hence 
\begin{equation*}
    P_i\big((\theta,b),\{(\theta',b'):\theta'\in A\}\big)\geq \frac{\rho\eta\int_{\mathbb{R}^d}\widetilde\pi_{i-1}(\gamma) L_i(\gamma)d\gamma}{\sup_{\gamma\in\mathbb{R}^d}\{L_i(\gamma)\}}  \widehat\Pi_i\big(\{(\theta',b'):\theta'\in A\}\big).
\end{equation*}
By this Doeblin condition (see, e.g., \citet{rosenthal1995minorization}), for any $t\in\mathbb{N}$, 
\begin{align*}
  \sup_{(\theta,b)\in\mathbb{R}^d\times [B]} d_{\mathrm{TV}}\left(P_i^t((\theta,b),\cdot),\widehat\Pi_i\right)
 \leq\,&\left(1-\frac{\rho\eta\int_{\mathbb{R}^d}\widetilde\pi_{i-1}(\gamma) L_i(\gamma)d\gamma}{\sup_{\gamma\in\mathbb{R}^d}\{L_i(\gamma)\}}\right)^t\nonumber\\
 \leq\,& \exp\left(-\frac{t\rho\eta\int_{\mathbb{R}^d}\widetilde\pi_{i-1}(\gamma) L_i(\gamma)d\gamma}{\sup_{\gamma\in\mathbb{R}^d}\{L_i(\gamma)\}}\right),
\end{align*}
from which the conclusion of the theorem follows.    

\subsection{Proof of Theorem~3.2}

Let $\widehat \pi_0:=\pi_0$. We prove the stronger claim that as $B\to\infty$, $\mathbb{E}[d_{\mathrm{TV}}(\widehat\pi_{i},\pi_{i})] \to 0$ for all $i\in \{0,1,\cdots,N\}$, and $\mathbb{E}[d_{\mathrm{TV}}(\widetilde\pi_{i-1},\pi_{i-1})]  \to 0$ for all $i\in \{1,\cdots,N\}$. 

We proceed by induction on $i\in\{0,1,\cdots,N\}$. Note that when $i=0$, $\mathbb{E}[d_{\mathrm{TV}}(\widehat\pi_0,\pi_0)]=0$. Below we consider $i\in [N]$, and assume that  
\begin{align}\label{con}
    \lim_{B\to\infty}\mathbb{E}[d_{\mathrm{TV}}(\widehat\pi_{i-1},\pi_{i-1})] = 0, \quad \text{and }\quad \lim_{B\to\infty}\mathbb{E}[d_{\mathrm{TV}}(\widetilde\pi_{i-2},\pi_{i-2})]=0 \text{ if } i\geq 2.
\end{align}

\noindent\textbf{Step 1.} We first show that 
\begin{align}\label{newq0}
    \lim_{B\to\infty}\mathbb{E}[d_{\mathrm{TV}}(\widetilde\pi_{i-1},\pi_{i-1})]= 0.
\end{align}

If $i=1$, by \cite[Theorem 9.2]{devroye2001combinatorial}, $\lim_{B\to\infty}\mathbb{E}[d_{\mathrm{TV}}(\widetilde\pi_0,\pi_0)]= 0$. Below we assume that $i\geq 2$. For any $\theta\in\mathbb{R}^d$, define
\begin{align*}
    p_{i-1,h}(\theta):=(K_h*\widehat\pi_{i-1})(\theta)=\int_{\mathbb{R}^d}K_h(\theta-\theta')\widehat\pi_{i-1}(\theta')d\theta'.
\end{align*}
By \cite[Theorem 9.1]{devroye2001combinatorial}, as $h\to 0$,  
\begin{equation*}
    \|K_h*\pi_{i-1}-\pi_{i-1}\|_1=\int_{\mathbb{R}^d}|K_h*\pi_{i-1}(\theta)-\pi_{i-1}(\theta)|d\theta\to 0.
\end{equation*}
Hence by~\eqref{con}, as $B\to\infty$, 
\begin{align}\label{e}
    \mathbb{E}[\|p_{i-1,h}-\pi_{i-1}\|_1]\leq\,& \mathbb{E}[\|K_h*\hat{\pi}_{i-1}-K_h*\pi_{i-1}\|_1]+\|K_h*\pi_{i-1}-\pi_{i-1}\|_1\nonumber\\
    \leq\,& \mathbb{E}[\|\widehat\pi_{i-1}-\pi_{i-1}\|_1]+\|K_h*\pi_{i-1}-\pi_{i-1}\|_1\nonumber\\
    =\,& 2\mathbb{E}[d_{\mathrm{TV}}(\widehat\pi_{i-1},\pi_{i-1})]+\|K_h*\pi_{i-1}-\pi_{i-1}\|_1\to 0.
\end{align}

Let $A_{i-2}$ be the event that $\int_{\mathbb{R}^d}\widetilde\pi_{i-2}(\theta)L_{i-1}(\theta)d\theta\geq \delta/2$. By Assumption~1 in the main paper and~\eqref{con}, 
\begin{align*}  \mathbb{E}\left[\left|\int_{\mathbb{R}^d}\widetilde\pi_{i-2}(\theta)L_{i-1}(\theta)d\theta-\int_{\mathbb{R}^d}\pi_{i-2}(\theta)L_{i-1}(\theta)d\theta\right|\right]\leq\,& \mathbb{E}\left[\int_{\mathbb{R}^d}|\widetilde\pi_{i-2}(\theta)-\pi_{i-2}(\theta)|L_{i-1}(\theta)d\theta\right]\nonumber\\
     \leq\,& 2M\mathbb{E}[d_{\mathrm{TV}}(\widetilde\pi_{i-2},\pi_{i-2})]\to 0. 
\end{align*}
As $\int_{\mathbb{R}^d}\pi_{i-2}(\theta)L_{i-1}(\theta)d\theta\geq \delta$, we have 
\begin{align}\label{c}
     \mathbb{P}(A_{i-2}^c)\leq \mathbb{P}\left(\left|\int_{\mathbb{R}^d}\widetilde\pi_{i-2}(\theta)L_{i-1}(\theta)d\theta-\int_{\mathbb{R}^d}\pi_{i-2}(\theta)L_{i-1}(\theta)d\theta\right|\geq \delta/2\right)\to 0.
\end{align}

Let $\mathcal{F}_{i-2}$ be the $\sigma$-algebra generated by $\{\theta_b^{(j)}\}_{b=1}^B$ for $0\leq j\leq i-2$. Let $\epsilon:=\min\big\{\frac{\rho\eta\delta}{2M},1\big\}\in (0,1]$. We have the following lemma.

\begin{lemma}\label{Lee1}
Suppose that $A_{i-2}$ holds. Then for any $R\geq 1$,  
\begin{align*}
    \mathbb{E}[\|\widetilde\pi_{i-1}-p_{i-1,h}\|_1\mid \mathcal{F}_{i-2}]\leq \sqrt{\frac{2\mathcal{L}^d(C_R)}{B\epsilon h^d} \int_{\mathbb{R}^d}K(u)^2du}+\frac{4}{\epsilon B}+2\int_{\mathbb{R}^d\backslash C_R}p_{i-1,h}(\theta)d\theta,
\end{align*}
where $\mathcal{L}^d$ is the Lebesgue measure on $\mathbb{R}^d$ and $C_R:=\{\theta\in\mathbb{R}^d:\|\theta\|_2\leq R\}$.
\end{lemma}
\begin{proof}

Throughout the proof, we condition on $\mathcal{F}_{i-2}$ and assume $A_{i-2}$ holds. Denote by $P_{i-1}$ the transition kernel of the graph-enabled Markov chain at stage $i-1$. From the proof of Theorem~3.3 and our assumption that $A_{i-2}$ holds, for any $b,b'\in [B]$, $\theta\in\mathbb{R}^d$, and any Borel measurable set $A\subseteq \mathbb{R}^d$, we have
\begin{equation}\label{Doeblin.condition}
    P_{i-1}\big((\theta,b),\{(\theta',b'):\theta'\in A\}\big)\geq \frac{\rho\eta\delta}{2M}\cdot  \widehat\Pi_{i-1}\big(\{(\theta',b'):\theta'\in A\}\big).
\end{equation}
From the above display, there exists a Markov kernel $R$ such that $P_{i-1}=\epsilon \widehat\Pi_{i-1}+(1-\epsilon)R$. Since $P_{i-1}$ preserves $\widehat\Pi_{i-1}$, $R$ preserves $\widehat\Pi_{i-1}$ as well. By induction, it can be shown that for any $t\in\mathbb{N}^*$, $P_{i-1}^t=(1-(1-\epsilon)^t)\widehat\Pi_{i-1}+(1-\epsilon)^t R^t$. For $t\in\mathbb{N}$, denote by $X_t=(\theta_t,b_t)$ the state of the graph-enabled Markov chain at step $t$, and by $\mu_t$ the distribution of $X_t$.

For any $\theta\in\mathbb{R}^d$, define 
\begin{equation*}
    \bar{p}_{i-1,h}(\theta):=\mathbb{E}[\widetilde\pi_{i-1}(\theta)]=\frac{1}{B}\sum_{t=1}^B (K_h*\nu_t)(\theta),
\end{equation*}
where $\nu_t$ is the distribution of $\theta_t$. By the Doeblin condition~\eqref{Doeblin.condition}, for any $t\in\mathbb{N}^*$,
\begin{equation*}
    d_{\mathrm{TV}}(\nu_t,\widehat\pi_{i-1})\leq d_{\mathrm{TV}}(\mu_t,\widehat\Pi_{i-1})\leq (1-\epsilon)^t.
\end{equation*}
Hence 
\begin{equation}\label{bddd}
    \|\bar{p}_{i-1,h}-p_{i-1,h}\|_1\leq \frac{2}{B}\sum_{t=1}^B d_{\mathrm{TV}}(\nu_t,\widehat\pi_{i-1})\leq \frac{2}{B}\sum_{t=1}^B(1-\epsilon)^t\leq \frac{2}{\epsilon B}.
\end{equation}

For any $\theta_0\in\mathbb{R}^d$ and $x=(\theta,b)\in\mathbb{R}^d\times [B]$, take $f_{\theta_0}(x):=K_h(\theta_0-\theta)$. Note that for any probability measure $\mu$ on $\mathbb{R}^d\times [B]$, we have
\begin{equation}\label{condd}
    \int_{\mathbb{R}^d} \mathbb{E}_{\mu}(f_{\theta_0}(X)^2)d\theta_0= \mathbb{E}_{\mu}\left[\int_{\mathbb{R}^d}K_h(\theta_0-\theta)^2d\theta_0\right]=h^{-d}\int_{\mathbb{R}^d}K(u)^2du.
\end{equation}

For $0\leq r<s$ and $f=f_{\theta_0}$ (where $\theta_0\in\mathbb{R}^d$),
\begin{equation*}
    \mathbb{E}[f(X_r)f(X_s)]=\mu_r[fP_{i-1}^{s-r}(f)]=(1-(1-\epsilon)^{s-r})\mu_r(f)\widehat\Pi_{i-1}(f)+(1-\epsilon)^{s-r}\mu_r[fR^{s-r}(f)]. 
\end{equation*}
Similarly, 
\begin{equation*}
    \mathbb{E}[f(X_s)]=(1-(1-\epsilon)^{s-r})\widehat\Pi_{i-1}(f)+(1-\epsilon)^{s-r}\mu_r[R^{s-r}(f)].
\end{equation*}
Hence by the Cauchy--Schwarz inequality and Jensen's inequality, we have
\begin{align}\label{condd1}
    \mathrm{Cov}(f(X_r),f(X_s))=\,&(1-\epsilon)^{s-r}\mathrm{Cov}_{\mu_r}(f,R^{s-r}(f))\nonumber\\
    \leq\,& (1-\epsilon)^{s-r}\mathbb{E}_{\mu_r}[f^2]^{1/2}\mathbb{E}_{\mu_r}[(R^{s-r}(f))^2]^{1/2}\nonumber\\
    \leq\,& (1-\epsilon)^{s-r}\mathbb{E}_{\mu_r}[f^2]^{1/2}\mathbb{E}_{\mu_r}[R^{s-r}(f^2)]^{1/2} \nonumber\\
    \leq \,&(1-\epsilon)^{s-r}\cdot\frac{\mathbb{E}_{\mu_r}[f^2]+\mathbb{E}_{\mu_r}[R^{s-r}(f^2)]}{2}.
\end{align}

By~\eqref{condd} and~\eqref{condd1}, 
\begin{equation*}
    \int_{\mathbb{R}^d}\mathrm{Cov}(f_{\theta_0}(X_r),f_{\theta_0}(X_s))d\theta_0\leq (1-\epsilon)^{s-r} h^{-d}\int_{\mathbb{R}^d}K(u)^2du.
\end{equation*}
Now $\widetilde\pi_{i-1}(\theta_0)=\frac{1}{B}\sum_{t=1}^Bf_{\theta_0}(X_t)$, hence
\begin{align*}
    \int_{\mathbb{R}^d}\mathrm{Var}\big(\widetilde\pi_{i-1}(\theta_0))d\theta_0=\,&\frac{1}{B^2}\sum_{r,s=1}^B \int_{\mathbb{R}^d}\mathrm{Cov}(f_{\theta_0}(X_r),f_{\theta_0}(X_s)) d\theta_0\nonumber\\
    \leq\,& \frac{1}{B^2}\sum_{r,s=1}^B(1-\epsilon)^{|r-s|} h^{-d}\int_{\mathbb{R}^d}K(u)^2du\leq \frac{2}{B\epsilon h^d} \int_{\mathbb{R}^d}K(u)^2du.
\end{align*}
Now fix any $R\geq 1$. By the Cauchy--Schwarz inequality, 
\begin{align*}
    \mathbb{E}\left[\int_{C_R}|\widetilde\pi_{i-1}(\theta)-\bar{p}_{i-1,h}(\theta)|d\theta\right]\leq\,& \mathcal{L}^d(C_R)^{1/2}\left(\int_{C_R}\mathrm{Var}(\widetilde\pi_{i-1}(\theta))  d\theta\right)^{1/2}\nonumber\\
    \leq\,& 
    \sqrt{\frac{2\mathcal{L}^d(C_R)}{B\epsilon h^d} \int_{\mathbb{R}^d}K(u)^2du}.
\end{align*}
Moreover, 
\begin{align*}
    \mathbb{E}\left[\int_{\mathbb{R}^d\backslash C_R}\widetilde\pi_{i-1}(\theta)d\theta\right]=\int_{\mathbb{R}^d\backslash C_R}\bar{p}_{i-1,h}(\theta)d\theta\leq \int_{\mathbb{R}^d\backslash C_R}p_{i-1,h}(\theta)d\theta+\|\bar{p}_{i-1,h}-p_{i-1,h}\|_1,
\end{align*}
\begin{align*}
    \|\widetilde\pi_{i-1}-p_{i-1,h}\|_1\leq\,& \int_{C_R}|\widetilde\pi_{i-1}(\theta)-\bar{p}_{i-1,h}(\theta)|d\theta+\int_{C_R}|\bar{p}_{i-1,h}(\theta)-p_{i-1,h}(\theta)|d\theta\nonumber\\
    &+\int_{\mathbb{R}^d\backslash C_R}\widetilde\pi_{i-1}(\theta)d\theta+\int_{\mathbb{R}^d\backslash C_R}p_{i-1,h}(\theta)d\theta.
\end{align*}
Combining the above three displays with~\eqref{bddd}, we conclude that
\begin{equation*}
    \mathbb{E}[\|\widetilde\pi_{i-1}-p_{i-1,h}\|_1]\leq \sqrt{\frac{2\mathcal{L}^d(C_R)}{B\epsilon h^d} \int_{\mathbb{R}^d}K(u)^2du}+\frac{4}{\epsilon B}+2\int_{\mathbb{R}^d\backslash C_R}p_{i-1,h}(\theta)d\theta.
\end{equation*}
\end{proof}

We have
\begin{equation*}
    \mathbb{E}\left[\int_{\mathbb{R}^d\backslash C_R}p_{i-1,h}(\theta)d\theta\right]\leq \int_{\mathbb{R}^d\backslash C_R}\pi_{i-1}(\theta)d\theta+\mathbb{E}[\|p_{i-1,h}-\pi_{i-1}\|_1].
\end{equation*}
Hence by~\eqref{e}, 
\begin{equation*}
    \lim_{B\to\infty}\mathbb{E}\left[\int_{\mathbb{R}^d\backslash C_R}p_{i-1,h}(\theta)d\theta\right]\leq \int_{\mathbb{R}^d\backslash C_R}\pi_{i-1}(\theta)d\theta.
\end{equation*}
By Lemma~\ref{Lee1}, for any $R\geq 1$,  
\begin{equation*}
    \mathbb{E}[\|\widetilde{\pi}_{i-1}-p_{i-1,h}\|_1]\leq \sqrt{\frac{2\mathcal{L}^d(C_R)}{B\epsilon h^d} \int_{\mathbb{R}^d}K(u)^2du}+\frac{4}{\epsilon B}+2  \mathbb{E}\left[\int_{\mathbb{R}^d\backslash C_R}p_{i-1,h}(\theta)d\theta\right]+2\mathbb{P}(A_{i-2}^c).
\end{equation*}
Taking first $B\to\infty$ and then $R\to\infty$, and noting~\eqref{c}, we obtain that 
\begin{equation}\label{cc}
    \lim_{B\to\infty}\mathbb{E}[\|\widetilde{\pi}_{i-1}-p_{i-1,h}\|_1]=0.
\end{equation}
The conclusion~\eqref{newq0} follows from~\eqref{e} and~\eqref{cc}.

\smallskip

\noindent\textbf{Step 2.} Now we show that 
\begin{align*}
    \lim_{B\to\infty}\mathbb{E}[d_{\mathrm{TV}}(\widehat\pi_{i},\pi_{i})] = 0.
\end{align*} 

For any $\theta\in\mathbb{R}^d$, we have
\begin{equation*}
    \widehat\pi_i(\theta)=\frac{\widetilde\pi_{i-1}(\theta)L_i(\theta)}{\int_{\mathbb{R}^d}\widetilde\pi_{i-1}(\gamma)L_i(\gamma)d\gamma},\qquad \pi_i(\theta)=\frac{\pi_{i-1}(\theta)L_i(\theta)}{\int_{\mathbb{R}^d}\pi_{i-1}(\gamma)L_i(\gamma)d\gamma}.
\end{equation*}
Hence 
\begin{align*}
    |\widehat\pi_i(\theta)-\pi_i(\theta)|\leq\, & \frac{\widetilde\pi_{i-1}(\theta)L_i(\theta)\int_{\mathbb{R}^d}|\widetilde\pi_{i-1}(\gamma)-\pi_{i-1}(\gamma)|L_i(\gamma)d\gamma}{\int_{\mathbb{R}^d}\widetilde\pi_{i-1}(\gamma)L_i(\gamma)d\gamma\int_{\mathbb{R}^d}\pi_{i-1}(\gamma)L_i(\gamma)d\gamma}\nonumber\\
    &+\frac{|\widetilde\pi_{i-1}(\theta)-\pi_{i-1}(\theta)|L_i(\theta)}{\int_{\mathbb{R}^d}\pi_{i-1}(\gamma)L_i(\gamma)d\gamma}.
\end{align*}
Therefore, we have  
\begin{align}\label{newq}
    \|\widehat\pi_i-\pi_i\|_1=\,&\int_{\mathbb{R}^d}|\widehat\pi_i(\theta)-\pi_i(\theta)|d\theta\leq \frac{2\int_{\mathbb{R}^d}|\widetilde\pi_{i-1}(\gamma)-\pi_{i-1}(\gamma)|L_i(\gamma) d\gamma}{\int_{\mathbb{R}^d}\pi_{i-1}(\gamma)L_i(\gamma)d\gamma}\nonumber\\
    \leq\,& 2M\delta^{-1}\int_{\mathbb{R}^d}|\tilde\pi_{i-1}(\gamma)-\pi_{i-1}(\gamma)|d\gamma   =  4M\delta^{-1}d_{\mathrm{TV}}(\widetilde\pi_{i-1},\pi_{i-1}),
\end{align}
where the second inequality uses Assumption~1 in the main paper. By~\eqref{newq0} and~\eqref{newq}, $\lim_{B\to\infty}\mathbb{E}[d_{\mathrm{TV}}(\widehat\pi_{i},\pi_{i})] = 0$.

\medskip

By \textbf{Steps 1} and \textbf{2}, the conclusion of the theorem follows by induction. 

\section{Additional Results for the OUD Simulation Study}\label{Appendix_B}

We assess the mixing behavior and statistical efficiency of FLaG-MCMC and the Metropolis random walk in the OUD simulation study in Section~4 of the main paper. Table~\ref{tab:oud-mixing} reports the MPSRFs and effective sample sizes; the effective sample sizes are averaged over three replications. The MPSRFs are close to one for both samplers across
$d=2,6,10$, indicating satisfactory mixing. FLaG-MCMC yields substantially
larger effective sample sizes,
approximately $2$--$3$ times those of the Metropolis random walk.

\begin{figure}[H]
     \centering
     \begin{subfigure}{0.9\textwidth}
         \centering
         \includegraphics[width=0.75\textwidth]{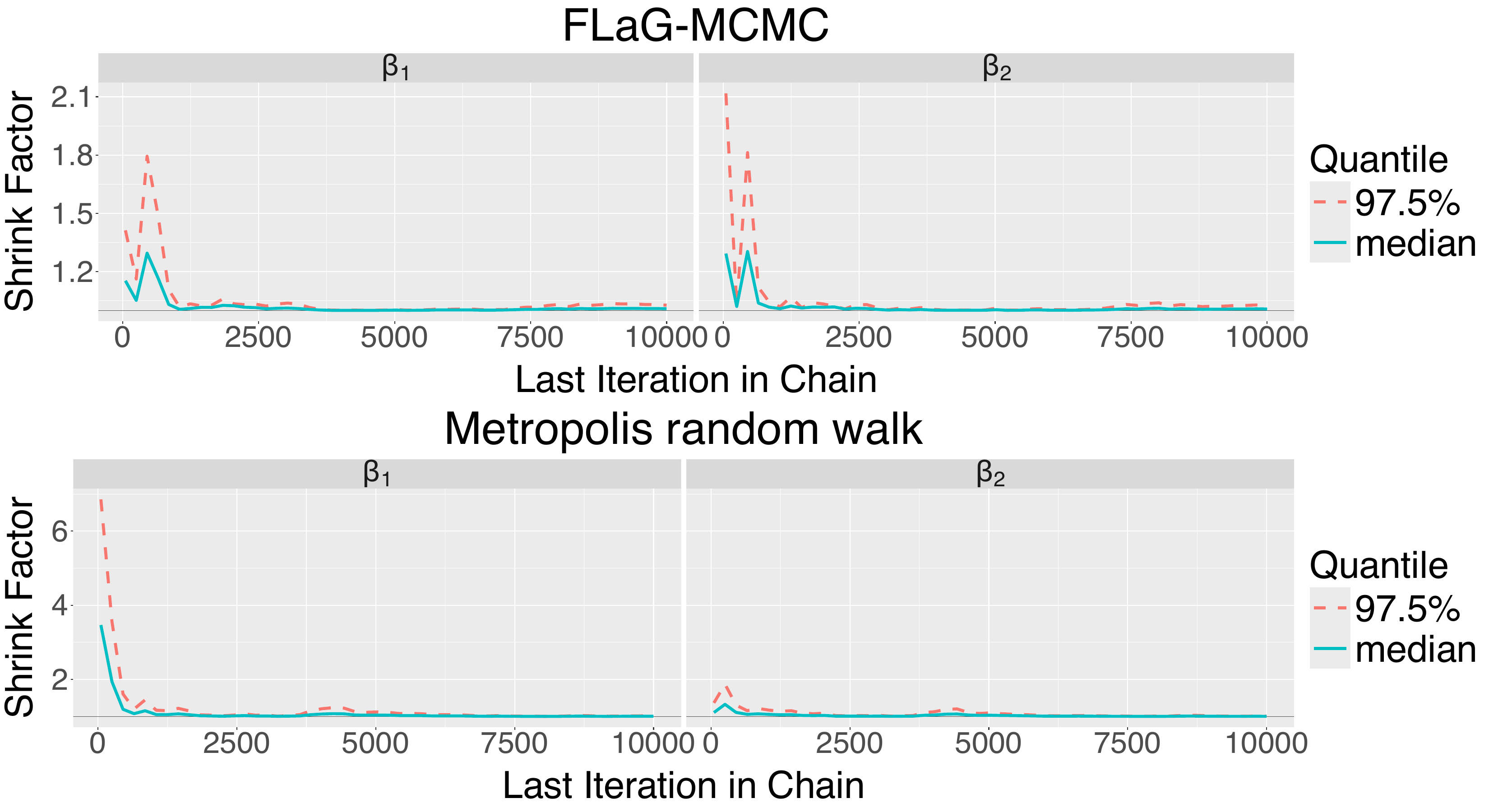}
         \caption{Gelman--Rubin--Brooks plots}
     \end{subfigure}
     \hfill
     \begin{subfigure}{0.9\textwidth}
         \centering
         \includegraphics[width=0.65\textwidth]{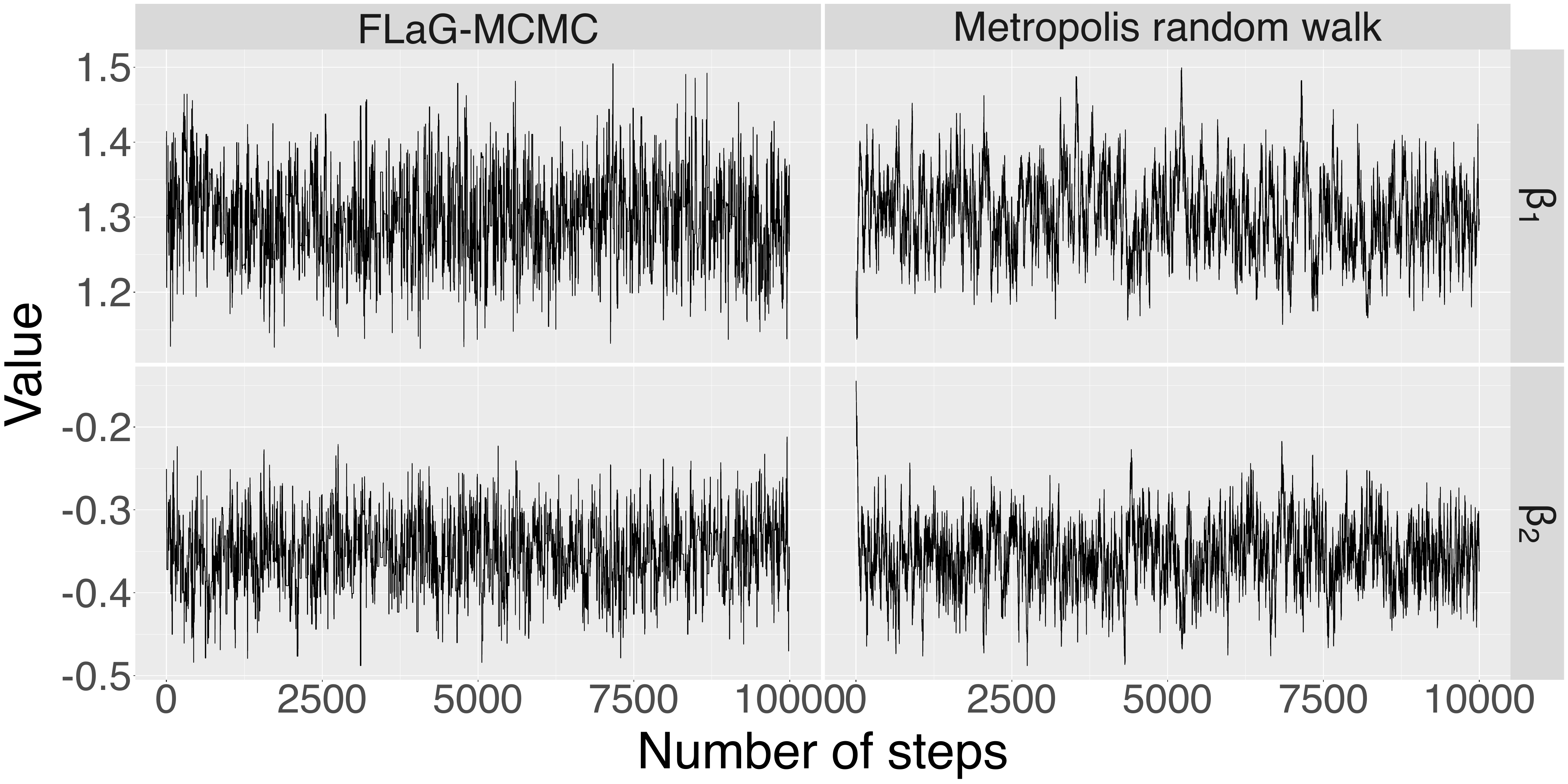}
         \caption{Traceplots}
     \end{subfigure}
     \hfill
     \begin{subfigure}{0.9\textwidth}
         \centering
         \includegraphics[width=0.6\textwidth]{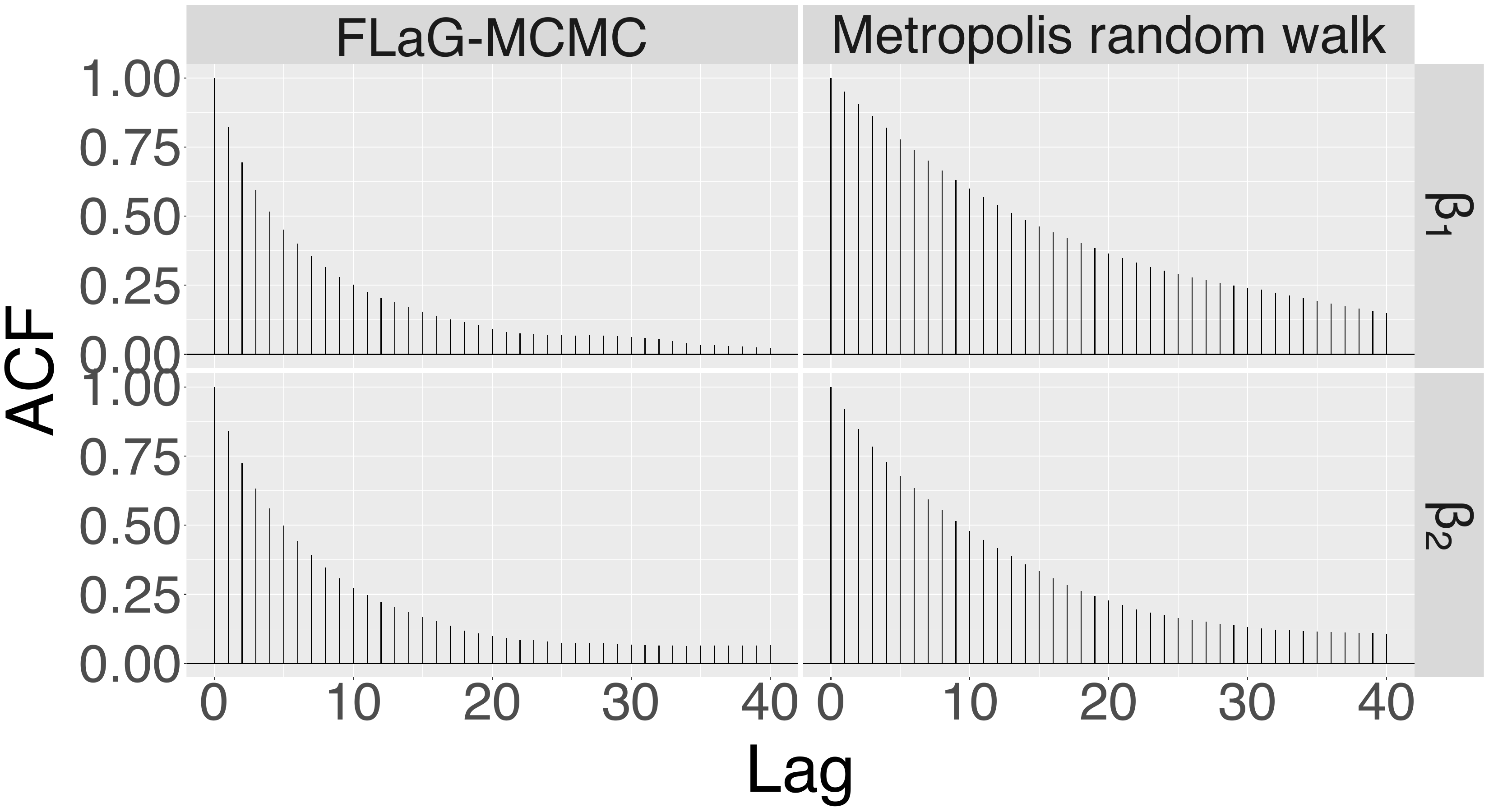}
         \caption{Autocorrelation plots}
     \end{subfigure}
        \caption{MCMC convergence diagnostics for the OUD simulation study with $d=2$}
        \label{Figure5.6(a)}
\end{figure}

\begin{figure}[H]
     \centering
     \begin{subfigure}{0.9\textwidth}
         \centering
         \includegraphics[width=0.75\textwidth]{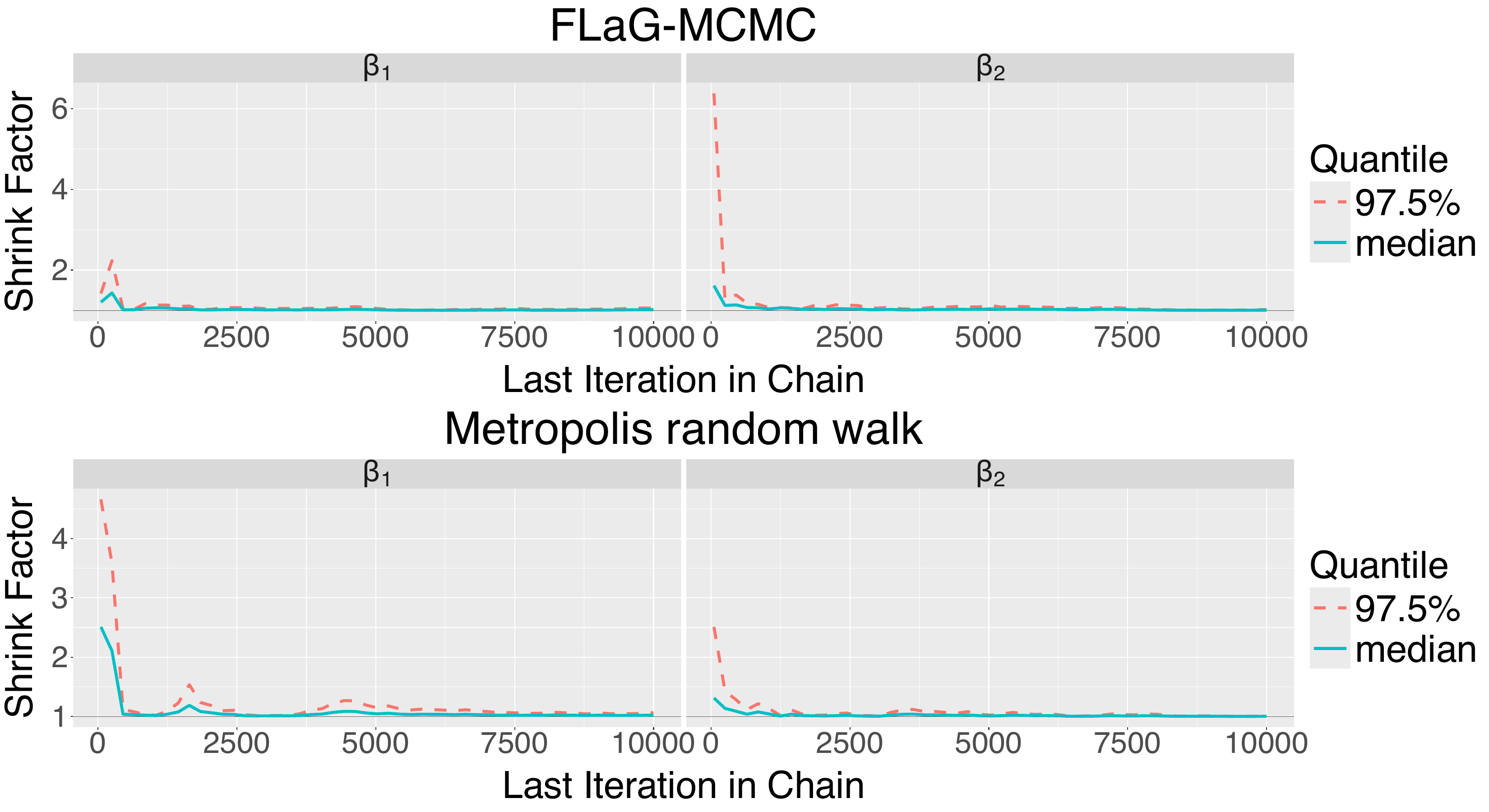}
         \caption{Gelman--Rubin--Brooks plots}
     \end{subfigure}
     \hfill
     \begin{subfigure}{0.9\textwidth}
         \centering
         \includegraphics[width=0.65\textwidth]{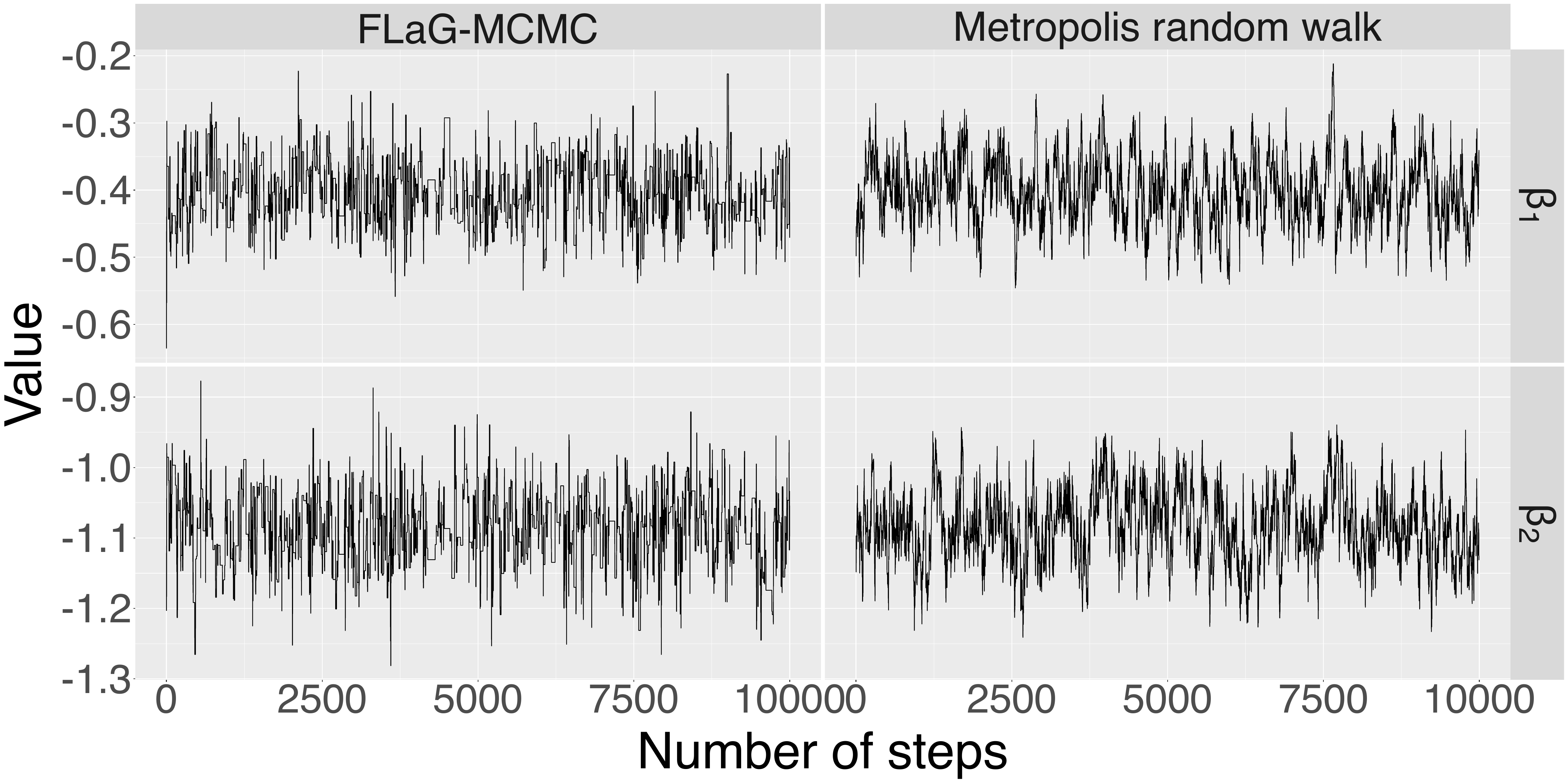}
         \caption{Traceplots}
     \end{subfigure}
     \hfill
     \begin{subfigure}{0.9\textwidth}
         \centering
         \includegraphics[width=0.6\textwidth]{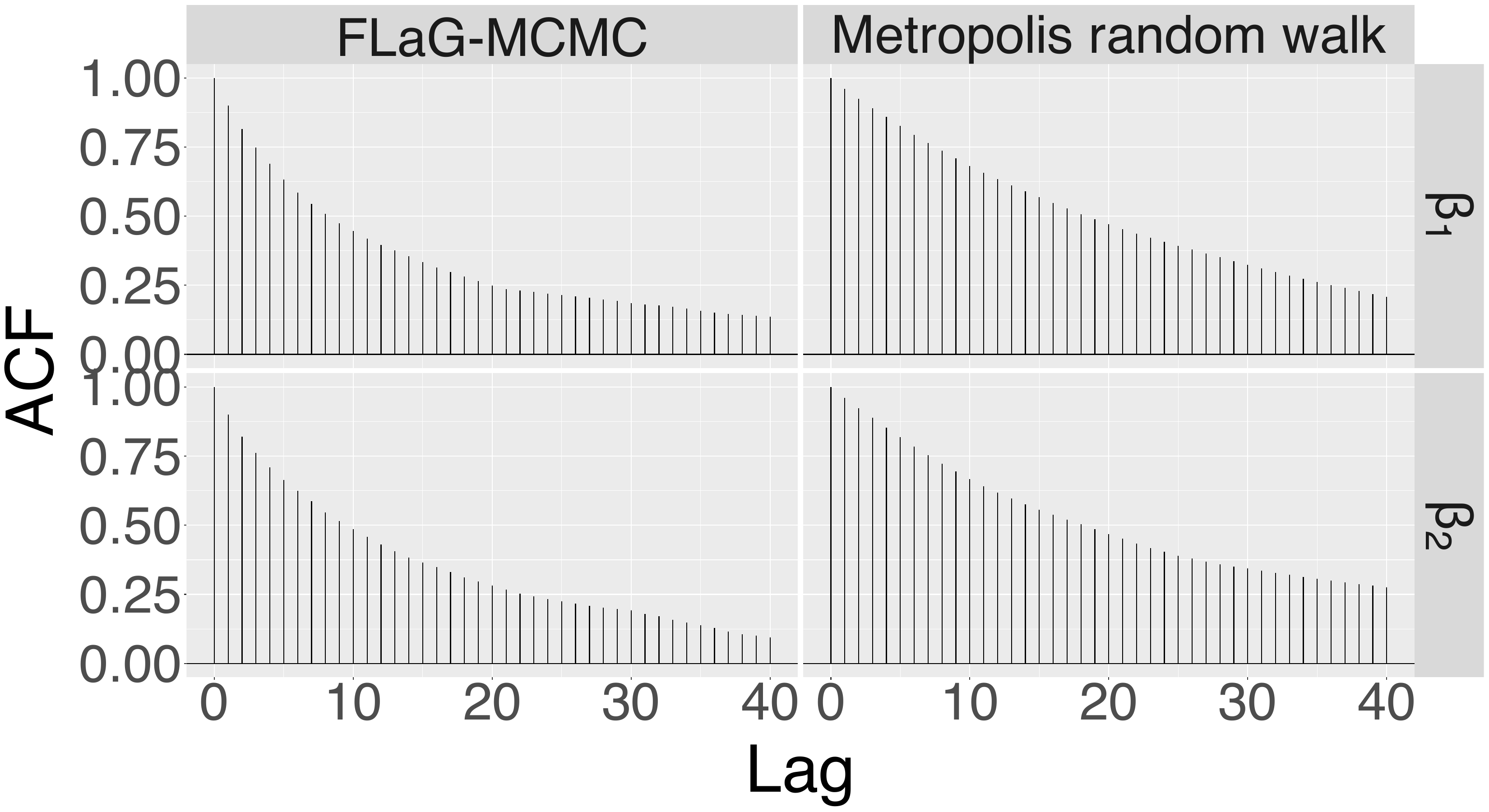}
         \caption{Autocorrelation plots}
     \end{subfigure}
        \caption{MCMC convergence diagnostics for the OUD simulation study with $d=6$}
        \label{Figure5.6(b)}
\end{figure}

\begin{figure}[H]
     \centering
     \begin{subfigure}{0.9\textwidth}
         \centering
         \includegraphics[width=0.75\textwidth]{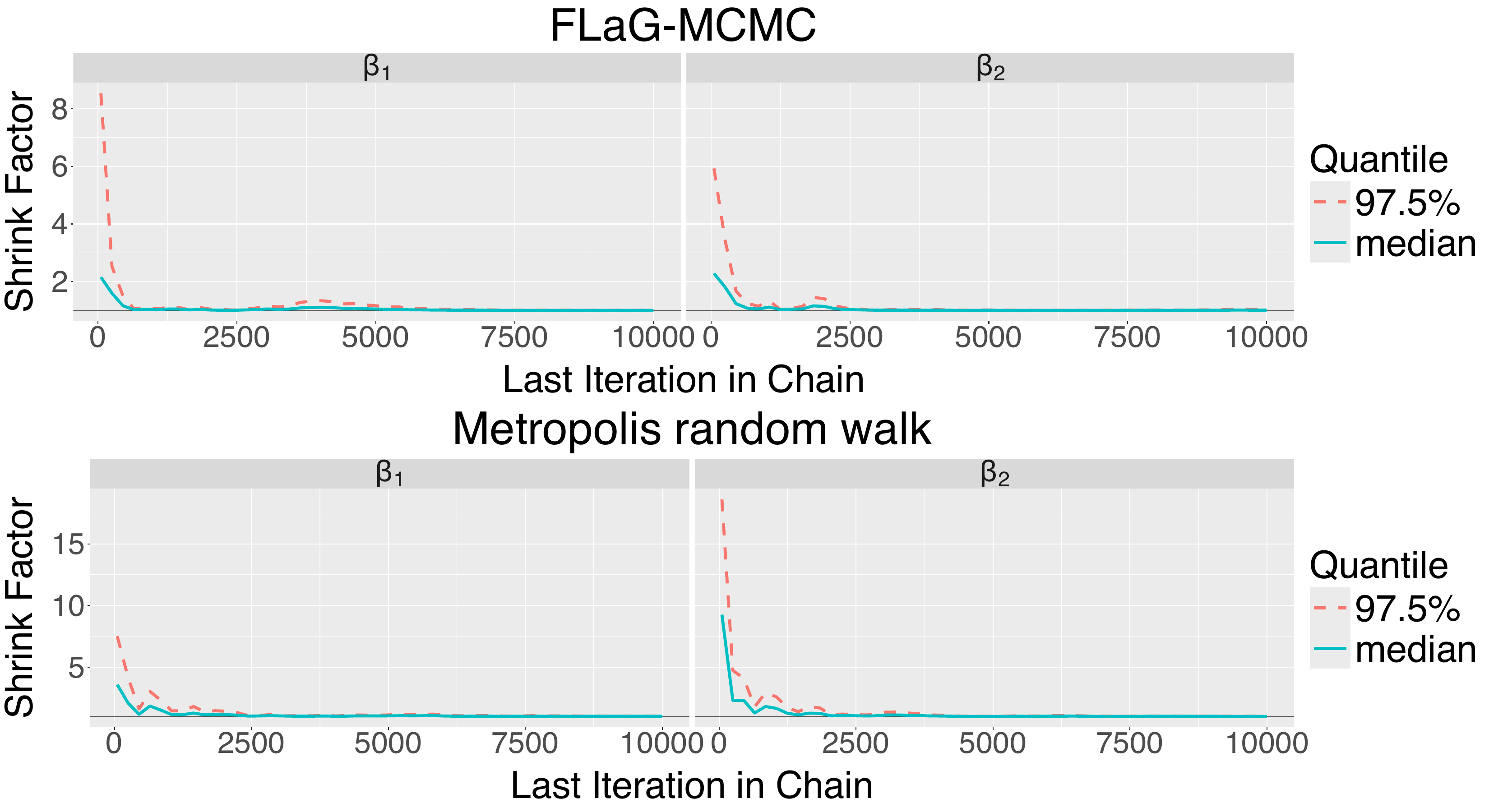}
         \caption{Gelman--Rubin--Brooks plots}
     \end{subfigure}
     \hfill
     \begin{subfigure}{0.9\textwidth}
         \centering
         \includegraphics[width=0.65\textwidth]{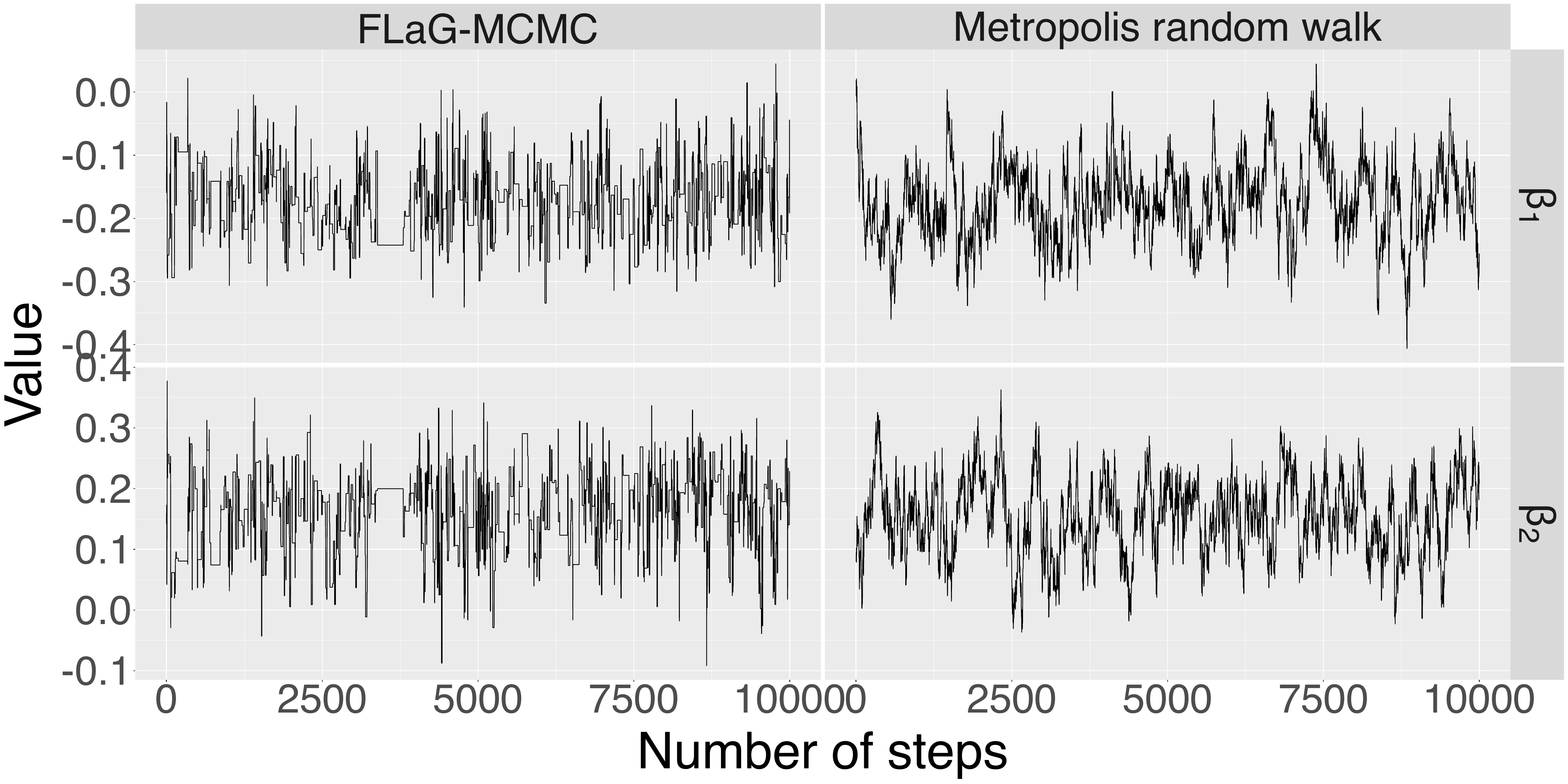}
         \caption{Traceplots}
     \end{subfigure}
     \hfill
     \begin{subfigure}{0.9\textwidth}
         \centering
         \includegraphics[width=0.6\textwidth]{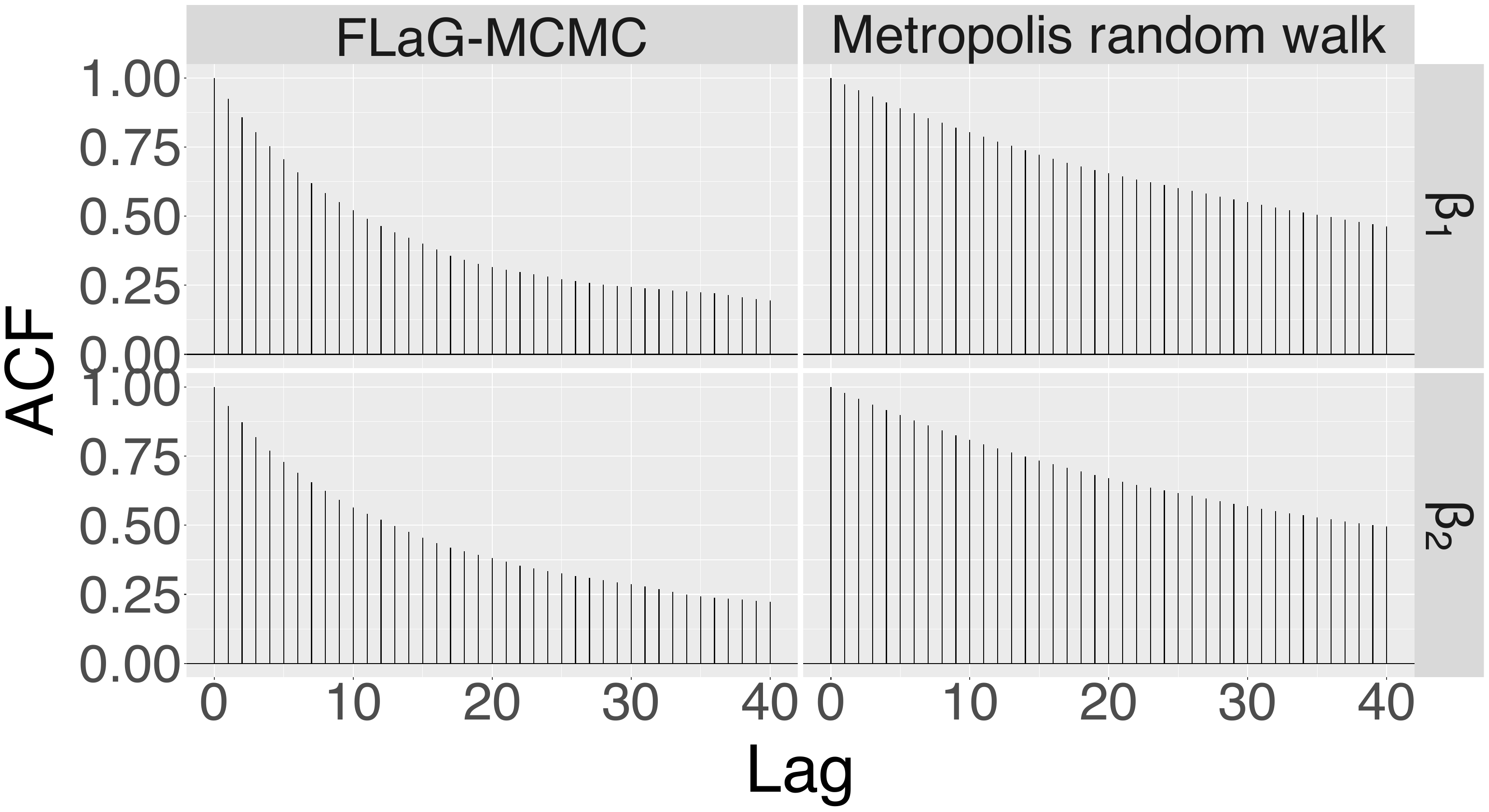}
         \caption{Autocorrelation plots}
     \end{subfigure}
        \caption{MCMC convergence diagnostics for the OUD simulation study with $d=10$}
        \label{Figure5.6(c)}
\end{figure}

Figures~\ref{Figure5.6(a)}--\ref{Figure5.6(c)} provide
the corresponding convergence and autocorrelation diagnostics. The
traceplots and Gelman--Rubin--Brooks plots support satisfactory mixing for
both samplers, while the autocorrelation plots show generally lower serial
dependence for FLaG-MCMC, consistent with its larger effective sample sizes.

\begin{table}
\caption{MCMC convergence and statistical efficiency for the federated OUD meta-analysis}
\label{tab:oud-mixing}
\centering
\begin{tabular}{@{}llccc@{}}
\hline
Metric & Method & $d=2$ & $d=6$ & $d=10$ \\
\hline
MPSRF
& FLaG-MCMC & 1.01 & 1.02 & 1.07 \\
& Metropolis random walk & 1.00 & 1.04 & 1.06 \\
\hline
Effective sample size, $\beta_1$
& FLaG-MCMC & 339 & 189 & 179 \\
& Metropolis random walk & 121 & 111 & 56 \\
\hline
Effective sample size, $\beta_2$
& FLaG-MCMC & 307 & 213 & 129 \\
& Metropolis random walk & 172 & 96 & 51 \\
\hline
\end{tabular}
\end{table}

\section{Additional Details for the GLOBEM Application}\label{Appendix_C}

The analysis in Section~5 of the main paper uses thirteen behavioral summaries derived from the location,
phone-use, physical activity, and sleep features released with GLOBEM
\citep{xu2022globem}. Table~\ref{tab:globem-feature-definitions} provides brief definitions of these behavioral summaries. For each depression assessment, each behavioral summary is computed by
averaging its available daily values over the strictly preceding 14-day window. The summaries are transformed using prespecified reference values and scales that are fixed before sequential fitting. 

\begin{figure}
\centering
\includegraphics[width=0.9\linewidth]{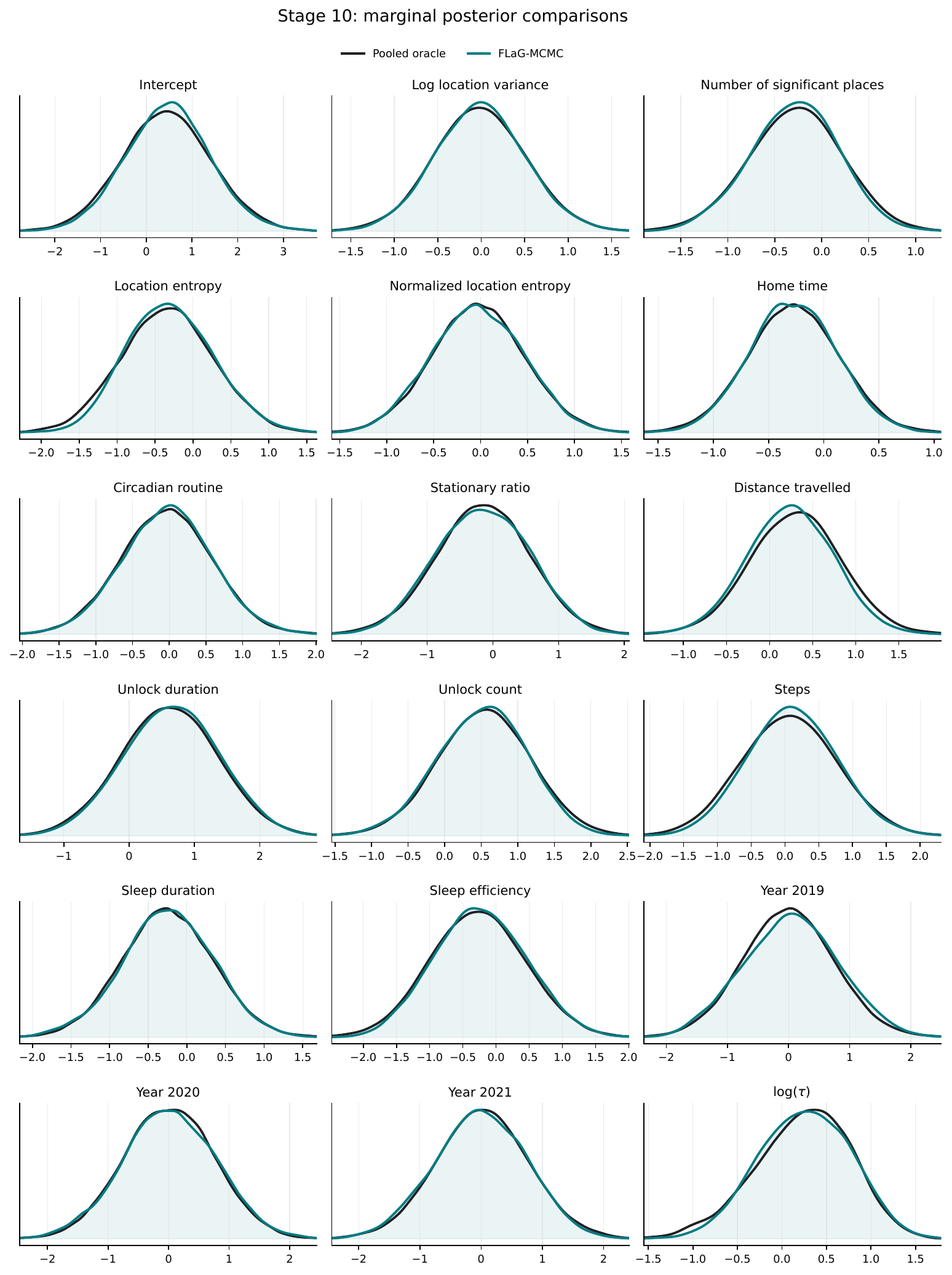}
\caption{Marginal posterior comparisons after the first $10$ participants
(2018). Black curves denote the pooled oracle and teal curves denote
FLaG-MCMC.}
\label{fig:stage10}
\end{figure}

\begin{figure}
\centering
\includegraphics[width=0.9\linewidth]{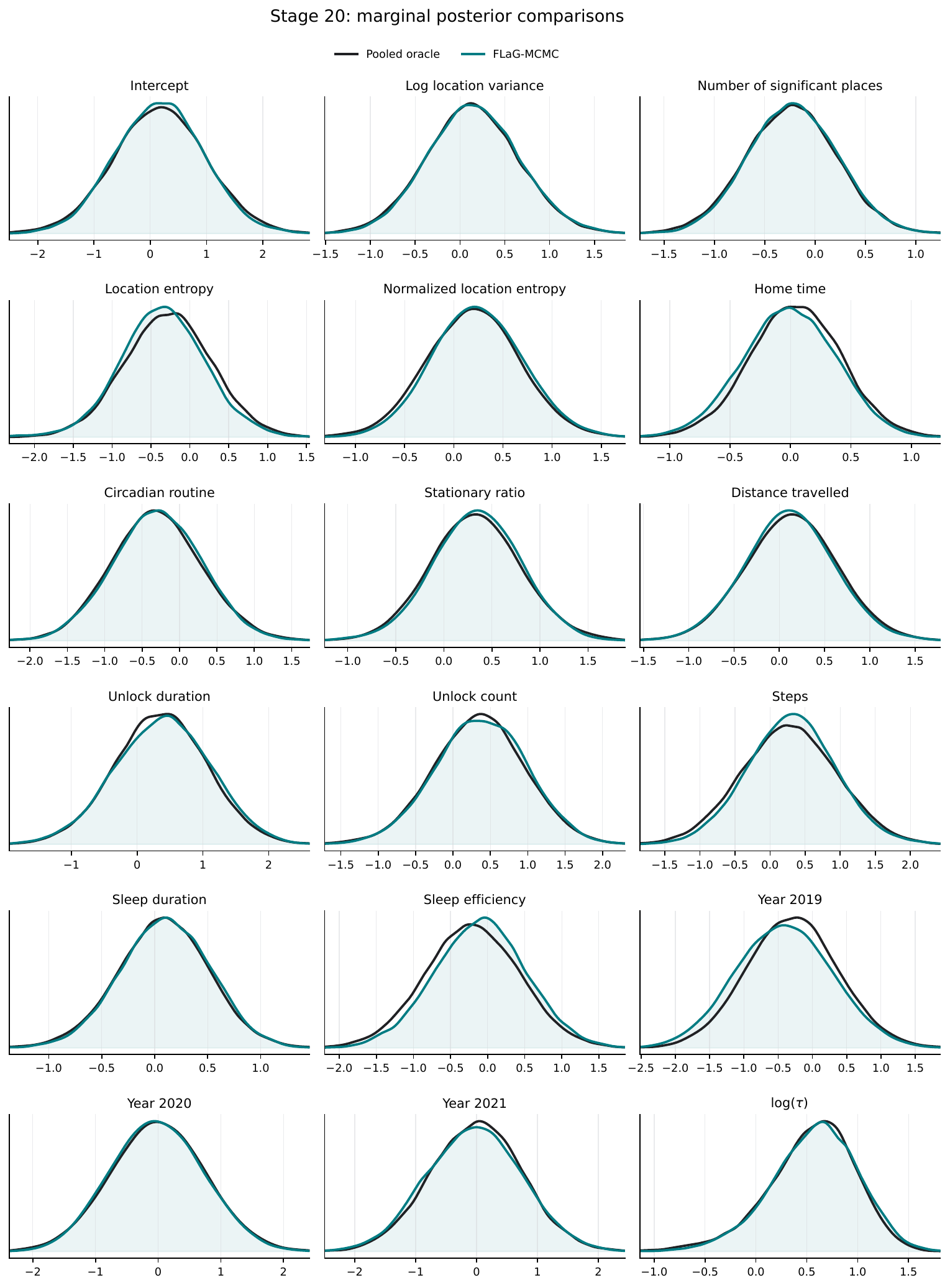}
\caption{Marginal posterior comparisons after $20$ participants (2018--2019). Black curves denote the pooled oracle and teal curves denote
FLaG-MCMC.}
\label{fig:stage20}
\end{figure}

\begin{figure}
\centering
\includegraphics[width=0.9\linewidth]{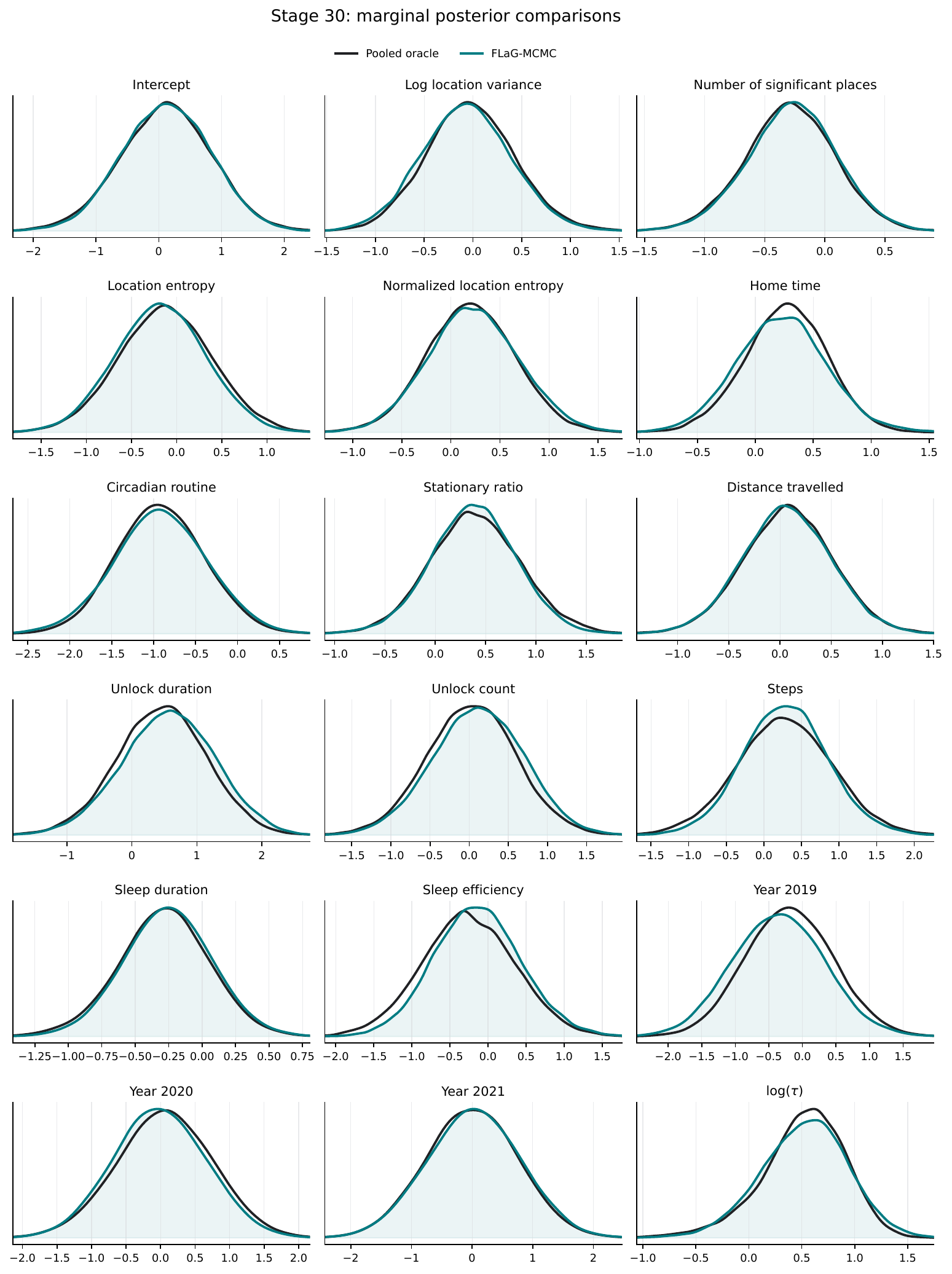}
\caption{Marginal posterior comparisons after $30$ participants (2018--2020). Black curves denote the pooled oracle and teal curves denote
FLaG-MCMC.}
\label{fig:stage30}
\end{figure}

For FLaG-MCMC, we use the moment-preserving shrinkage representation in Section~2.2.2, specifically the whitening-based construction with $\widetilde{K}(u)=(2\pi)^{-d/2}\exp(-\|u\|^2/2)$ and $h=0.22$. For graph construction, the posterior reservoir is projected onto a $5$-dimensional random subspace, after which we construct a $k$-nearest-neighbor graph with $k=32$. We set the global refresh
probability to $\rho=0.05$. Marginal posterior comparisons between FLaG-MCMC and the pooled oracle at intermediate stages 10,
20, and 30 are presented in Figures~\ref{fig:stage10}--\ref{fig:stage30}.

\begin{table}[htbp]
\centering
\caption{Behavioral summaries used in the GLOBEM application}
\label{tab:globem-feature-definitions}
\small
\renewcommand{\arraystretch}{1.15}
\begin{tabularx}{\textwidth}{
    >{\raggedright\arraybackslash}p{0.25\textwidth}
    >{\raggedright\arraybackslash}X
}
\toprule
Behavioral summary & Definition \\
\midrule

Log location variance
& Base-10 logarithm of the sum of the variances of the latitude and longitude coordinates \\

Number of significant places
& Number of significant location clusters represented in the daily segment \\

Location entropy
& Shannon entropy computed from the row-count distribution across significant location clusters \\

Normalized location entropy
& Location entropy divided by the number of significant location clusters represented in the segment \\

Home time
& Time spent at home, where home is defined as the most visited significant location between 8~p.m. and 8~a.m., including pauses within 200 meters of that location \\

Circadian routine
& Measure of similarity between a participant's daily mobility pattern and patterns observed on other sensed days, ranging from 0 to 1, with larger values indicating greater regularity \\

Stationary ratio
& Proportion of sensed location time classified as stationary, where a coordinate interval is classified as stationary when its estimated speed to the next coordinate is below 1 km/hour \\

Distance travelled
& Total distance traversed during movement (``flight'') segments in the pause--flight representation of the daily mobility trajectory \\

Unlock duration
& Total duration of phone-unlock episodes in the daily segment \\

Unlock count
& Number of phone-unlock episodes in the daily segment \\

Steps
& Average daily step count in the corresponding daily segment \\

Sleep duration
& Average duration asleep among sleep records designated as main sleep \\

Sleep efficiency
& Average reported sleep efficiency among records designated as main sleep \\

\bottomrule
\end{tabularx}
\end{table}

\end{document}